\documentclass[letterpaper,twocolumn,10pt]{article}
\usepackage{usenix}

\usepackage{tikz}
\usepackage{amsmath}

\usepackage{filecontents}

\usepackage{listings}
\usepackage{amssymb}
\usepackage{pifont}
\usepackage{graphicx}
\usepackage{todonotes}
\usepackage{comment}
\usepackage[toc,page]{appendix}

\usepackage{algorithm}
\usepackage{algpseudocode}

\relax
\relax

\lstdefinelanguage
[x64]{Assembler}
[x86masm]{Assembler}
{morekeywords={CDQE,CQO,CMPSQ,CMPXCHG16B,JRCXZ,LODSQ,MOVSXD,
POPFQ,PUSHFQ,SCASQ,STOSQ,IRETQ,RDTSCP,SWAPGS,
CLFLUSH,CMOVE,CMOVNE,MOVQ,CPUID,MFENCE
rax,rdx,rcx,rbx,rsi,rdi,rsp,rbp,
r8,r8d,r8w,r8b,r9,r9d,r9w,r9b,
r10,r10d,r10w,r10b,r11,r11d,r11w,r11b,
r12,r12d,r12w,r12b,r13,r13d,r13w,r13b,
r14,r14d,r14w,r14b,r15,r15d,r15w,r15b}}

\lstdefinelanguage{C2}[]{C}{
morekeywords={size_t,bool,true}
}

\newcommand{\etal}{\textit{et al.}}
\newcommand{\ww}{\textsc{Write+Write}}
\newcommand{\cmark}{\ding{51}}\newcommand{\xmark}{\ding{55}}

\usepackage{standalone}
\usepackage{caption}
\usepackage{pgfplots}
\begin{document}

\date{}

\title{\Large \bf Write Me and I'll Tell You Secrets -- \\ Write-After-Write Effects On Intel CPUs}

\author{
{\rm
Jan
Philipp
Thoma}\\
Ruhr-University
Bochum\\
jan.thoma@rub.de
\and
{\rm
Tim
Güneysu}\\
Ruhr-University
Bochum\\
tim.gueneysu@rub.de
}
\maketitle\begin{abstract}
There
is
a
long
history
of
side
channels
in
the
memory
hierarchy
of
modern
CPUs.
Especially
the
cache
side
channel
is
widely
used
in
the
context
of
transient
execution
attacks
and
covert
channels.
Therefore,
many
secure
cache
architectures
have
been
proposed.
Most
of
these
architectures
aim
to
make
the
construction
of
eviction
sets
infeasible
by
randomizing
the
address-to-cache
mapping.

In
this
paper,
we
investigate
the
peculiarities
of
write
instructions
in
recent
CPUs.
We
identify
\ww{},
a
new
side
channel
on
Intel
CPUs
that
leaks
whether
two
addresses
contend
for
the
same
cache
set.
We
show
how
\ww{} can be used for rapid construction of eviction sets on current cache architectures. Moreover,
we
replicate
the
\ww{}
effect
in
gem5
and
demonstrate
on
the
example
of
ScatterCache~\cite{DBLP:conf/uss/WernerUG0GM19}
how
it
can
be
exploited
to
efficiently
attack
state-of-the-art
cache
randomization
schemes.
In
addition
to
the
\ww{}
side
channel,
we
show
how
Write-After-Write
effects
can
be
leveraged
to
efficiently
synchronize
covert
channel
communication
across
CPU
cores.
This
yields
the
potential
for
much
more
stealthy
covert
channel
communication
than
before.
\end{abstract}

\section{Introduction}
It
is
no
secret
that
the
microarchitecture
of
recent
CPUs
is
riddled
with
side
channels
that
can
often
be
exploited
in
ways
that
threaten
the
security
of
the
whole
system.
Many
of
these
side
channels
are
the
result
of
obvious
and
elementary
CPU
components
that
pave
the
way
to
achieve
the
performance
levels
to
which
we
have
become
so
accustomed.
Among
others,
this
includes
caches
and
prefetchers.
The
way
these
components
are
intended
to
work
generates
a
timing
difference
that
can
be
observed
by
user-level
processes.
On
the
other
hand,
there
are
more
subtle
aspects
of
CPU
internals
and
their
implementation
that
lead
to
measurable
timing
differences
without
being
essential
for
CPUs
performance
or
security.
This,
for
example,
includes
Intel's
ring
interconnect
implementation
for
last
level
caches
(LLCs)~\cite{DBLP:conf/uss/PaccagnellaLF21}
or
store-to-load
forwarding~\cite{DBLP:journals/corr/abs-1905-05725}.

Due
to
the
vast
performance
discrepancy
between
the
CPU
core
and
the
memory
subsystem,
the
read-
and
write
path's
are
subject
to
immense
optimization
efforts
by
the
CPU
developers.
Each
saved
or
predicted
interaction
with
the
memory
can
result
in
hundreds
of
saved
clock
cycles.
Though
the
performance
benefit
of
such
optimizations
stands
without
question,
ongoing
research
has
found
many
ways
to
exploit
these
to
bypass
essential
security
foundations.
Early
work
in
this
area
demonstrated
how
cache-timing
can
be
used
to
reconstruct
secret
keys
of
AES~\cite{bernstein2005cache,Osvik-2006-CacheAttacksandCo}.
Over
time,
these
attacks
developed
towards
well-known
attack-primitives
like
\textsc{Prime+Probe}~\cite{DBLP:journals/joc/TromerOS10,Osvik-2006-CacheAttacksandCo,Liu-2015-Last-LevelCacheSid}
and
\textsc{Flush+Reload}~\cite{DBLP:conf/uss/YaromF14}.
With
these
primitives,
cache
attacks
evolved
to
be
very
efficient
and
further
CPU
components
like
the
TLB
moved
into
focus~\cite{DBLP:conf/uss/GrasRBG18}.
In
2018,
the
disclosure
of
Meltdown~\cite{Lipp2018meltdown}
and
Spectre~\cite{Kocher2018spectre}
shifted
the
momentum
and
severity
of
microarchitectural
attacks.
The
following
avalanche
of
transient
execution
attacks
changed
the
understanding
of
the
hardware
as
a
trust
anchor
for
secure
system
development;
see
generally~\cite{canella2019systematic}.
The
class
of
transient
execution
attacks
goes
beyond
control
flow
speculation,
i.e.
by
the
branch
predictor.
For
example,
the
MDS
attacks~\cite{DBLP:conf/sp/SchaikMOFMRBG19,DBLP:conf/ccs/CanellaGGGLMMP019}
exploit
speculative
data
forwarding
of
read-
and
write
operations.

During
transient
execution
attacks,
leaked
data
is
usually
recovered
via
a
covert
channel.
Thereby,
the
attacker
transmits
data
from
the
(speculative)
victim
context
to
their
own
process
using
timing
peculiarities
of
CPU
internals.
Covert
channels
can
also
be
used
to
communicate
between
co-located
VMs
in
cloud
environments~\cite{DBLP:conf/ccs/RistenpartTSS09}.
Due
to
the
simplicity
and
reliability
of
cache
covert
channels,
\textsc{Flush+Reload}, \textsc{Prime+Probe} and
derivatives\cite{DBLP:conf/dimva/GrussMWM16,DBLP:conf/ccs/PurnalTV21}
are
commonly
used
in
this
context.
The
bandwidth
of
such
covert
channels
has
shown
to
be
more
than
sufficient
to
transmit
large
chunks
of
data~\cite{maurice2017hello}.
However,
synchronization
across
cores
remains
an
issue
and
is
frequently
evaded
by
using
self-clocking
signals~\cite{tanenbaum1996computer}
with
massive
oversampling
on
the
receiver
end
and
multiple
accesses
on
the
sender
side~\cite{DBLP:conf/uss/WuXW12,maurice2017hello}.

\paragraph{Contributions.} In this paper, we  present \ww{}, a new
write-based
side
channel
on
Intel
CPUs
that
leaks
whether
two
physical
addresses
collide
in
a
specific
range.
This
side
channel
is
especially
worrisome
in
face
of
the
current
development
in
cache
side
channel
countermeasures.
We
replicate
the
behavior
in
\textit{gem5}~\cite{DBLP:journals/corr/abs-2007-03152} and demonstrate
an
improved
attack
against
state-of-the-art
cache
randomization
on
the
example
of
ScatterCache~\cite{DBLP:conf/uss/WernerUG0GM19}.
Our
attack
requires
further
design
constraints
to
be
considered
when
implementing
randomized
caches.
Secondly,
we
show
how
\ww{}
affects
traditional
cache
architectures
and
leverage
the
side
channel
for
bottom-up
construction
of
cache
eviction
sets.
In
doing
so,
we
break
the
current
speed
records
in
eviction
set
construction.
Third,
we
present
a
new,
write-based
technique,
to
synchronize
processes
across
CPU
cores.
We
show
how
this
technique
can
be
applied
to
establish
a
common
clock
signal
for
covert
channels.
Using
our
synchronization
approach,
each
signal
only
needs
to
be
transmitted
once
which
greatly
reduces
the
monitoring
surface
for
detection
mechanisms.

\textit{A version of this paper was sent to Intel for responsible disclosure
prior
to
submission
to
RAID'22.
Proof-of-concept
code
is
available
on
GitHub\footnote{\url{https://github.com/Chair-for-Security-Engineering/Write-Write}}.}

\paragraph{Organization of this Paper.} The following section introduces
background
on
caches,
cache
side-
and
covert
channels,
as
well
as
some
internals
of
recent
x86
CPUs.
In
\autoref{write+write:sec:write_channel},
we
introduce
the
\ww{}
side
channel
and
the
foundation
for
our
synchronization
technique.
We
then
present
a
\ww
-based
algorithm
for
rapid
eviction
set
construction
in
\autoref{write+write:sec:evsets}.
In
\autoref{write+write:sec:evsets_rand},
we
adapt
the
algorithm
for
randomized
caches
and
attack
a
gem5
implementation
of
ScatterCache.
Third,
we
demonstrate
the
Write-After-Write-based
cross-core
synchronization
for
covert
channel
communication
in
\autoref{write+write:sec:sync}. We discuss mitigation techniques and
related
work
in
\autoref{write+write:sec:mitigation}
and
\autoref{write+write:sec:related}
respectively.
Finally,
we
conclude
in
\autoref{write+write:sec:conclusion}.
\section{Background}
In
this
section,
we
introduce
some
background
on
caches,
covert
channels,
and
the
x86
microarchitecture.

\subsection{Caches}
The
speed
at
which
modern
processors
execute
instructions
greatly
exceeds
the
speed
of
read
and
write
operations
from
and
to
the
memory.
Since
many
programs
rely
on
frequent
memory
accesses,
this
would
normally
cause
a
large
number
of
stall
cycles,
waiting
for
the
requested
data
to
be
fetched.
Hence,
apart
from
deeply
embedded
devices,
virtually
all
current
processors
feature
at
least
one
level
of
cache.

Caches
are
small
and
fast
memory
modules
located
in
close
physical
proximity
to
the
CPU.
Frequently
used
data
is
stored
in
the
cache
to
accelerate
memory
operations
and
hide
the
latency
of
the
main
memory.
Most
desktop-level
processors
feature
three
levels
of
cache.
The
L1-cache
is
the
smallest
and
fastest
cache,
followed
by
the
slightly
larger
and
slower
L2-cache.
Both
L1-
and
L2-caches
are
typically
duplicated
for
each
physical
CPU
core.
The
last-level-cache
(LLC)
is
the
largest
level
of
cache
and
usually
shared
among
cores.
A
coherency
protocol
is
implemented
to
keep
the
data
consistent
across
all
caches
and
the
main
memory;
for
details
on
recent
Intel
CPUs,
see~\cite{DBLP:conf/icpp/MolkaHSN15}.
Furthermore,
the
LLC
is
usually
inclusive
which
means
that
all
entries
of
the
L1
and
L2
caches
are
also
stored
in
the
LLC.
This
brings
performance
benefits
in
multi-core
systems
-
if
a
L2
cache
miss
occurs,
the
inclusiveness
makes
sure
that
\textit{if}
the
data
is
cached
in
any
other
cores
private
cache,
it
is
also
cached
in
the
LLC.
Non-inclusive
LLCs
need
to
query
other
cores'
private
caches
or
maintain
a
directory~\cite{DBLP:conf/sp/YanSGFCT19}
to
make
sure
that
these
do
not
hold
a
modified
copy
of
the
requested
data.

\begin{figure}
\centering
\includegraphics[width=\linewidth]{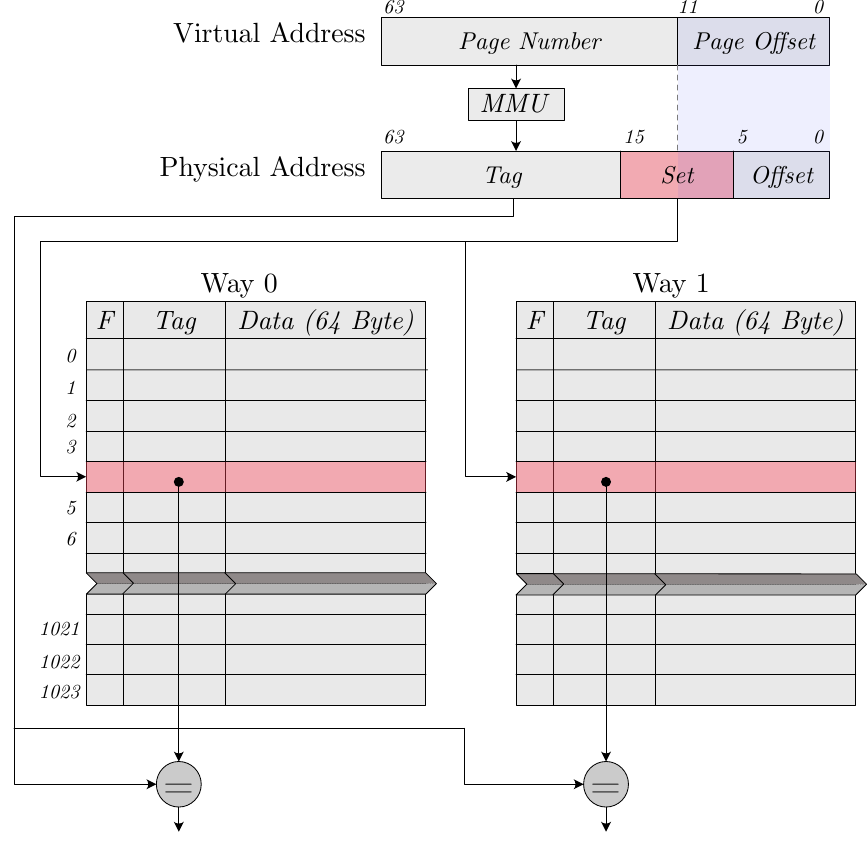}
\caption{\label{write+write:fig:cache_overview}Exemplary architecture of a physically indexed two-way set-associative cache. The index bits of the physical address are used to determine the set (red). The replacement policy chooses which entry is replaced on a miss access.}
\end{figure}

\paragraph{Cache Internals \& Addressing.}
Since
low
latency
is
a
key
design
goal
of
caches,
it
is
not
practical
to
search
the
whole
cache
on
every
access.
To
accelerate
the
lookup,
caches
are
usually
implemented
as
set-associative
structures.
Each
entry
(\textit{cache
line})
holds
64
bytes
of
data
alongside
a
tag,
which
is
used
to
uniquely
identify
the
cached
address,
and
some
flags
including
valid
and
dirty.
As
depicted
in
\autoref{write+write:fig:cache_overview}, the physical address is divided
into
\textit{tag}-,
\textit{set}-
and
\textit{offset}-bits.
The
offset
is
used
to
select
a
64-bit
word
from
the
cache
line
to
be
returned
on
read-access.
The
set-bits
select
the
cache
set
(corresponding
to
the
table
row
in
each
cache
way
in
\autoref{write+write:fig:cache_overview}).
The
remainder
of
the
address
(i.e.
the
tag)
is
stored
alongside
the
data
which
together
with
the
implicitly
stored
set
index,
uniquely
identifies
the
physical
address.

When
a
memory
address
is
accessed,
a
cache
lookup
occurs
and
in
each
cache
way,
the
tag
stored
at
the
index
determined
by
the
set
bits
of
the
address
are
compared
to
the
tag
of
the
accessed
address.
If
the
tags
match
in
one
cache
way,
a
hit
occurs
and
the
data
is
returned
with
the
specified
offset.
If
the
tags
do
not
match
in
any
cache
way,
a
cache
miss
occurs
and
the
data
is
requested
from
the
next
device
in
the
memory
hierarchy.
When
the
request
is
served,
the
\textit{replacement
policy}
selects
one
of
the
set
entries
to
be
replaced
with
the
new
data.
Often,
this
policy
is
(pseudo)-least-recently-used
((P)LRU)
which
replaces
the
entry
that
has
not
been
used
longest.
Writes
are
handled
analogously,
although
a
distinction
is
made
between
write-back
and
write-through
caches.
Write-back
caches
store
a
modified
version
of
the
data
until
the
entry
gets
evicted
from
the
cache
while
write-through
caches
immediately
forward
the
modification
to
the
memory-side
port.
Recent
LLCs
are
usually
configured
as
write-back.

In
addition
to
that,
many
processors
use
cache
slices
which
can
be
imagined
as
load-balanced,
parallel
instantiations
of
caches
to
reduce
the
workload
on
each
slice
and
increase
the
overall
bandwidth
of
the
cache.
Each
physical
address
is
uniquely
mapped
to
a
single
slice.
Recent
Intel
processors
implement
complex
cache
indexing
which
derives
the
cache
slice
by
using
a
recently
revealed
function
that
operates
on
``potentially
all''
address
bits~\cite{Intel2020}.
In~\cite{DBLP:conf/ndss/HundWH13},
this
complex
addressing
function
was
first
reverse
engineered
manually,
followed
by~\cite{DBLP:conf/raid/MauriceSNHF15}
which
utilizes
a
generic
method
based
on
hardware
performance
counters
to
reverse
engineer
the
function
for
several
Intel
processors.
Both
works
report
a
simple
xor-based
function
to
obtain
the
slice
for
each
address.

\paragraph{Cache Side Channels.}
The
design
goal
of
caches
is
to
accelerate
slow
memory
accesses
-
hence,
the
fact
that
timing
measurements
can
reveal
whether
or
not
some
data
was
cached
is
conceptually
unavoidable.
An
attacker
can
measure
the
latency
of
a
memory
access
and
therefore
determine
whether
the
accessed
data
was
cached
before
the
access.
This
effect
has
been
exploited
for
numerous
attacks
including
key-recovery
on
cryptographic
schemes~\cite{bernstein2005cache,DBLP:conf/eurosec/GotzfriedESM17},
bypassing
ASLR~\cite{DBLP:conf/ndss/GrasRBBG17,DBLP:conf/ndss/HundWH13},
covert
channels
in
shared
cloud
environments~\cite{maurice2017hello},
and
in
the
context
of
speculative
execution
attacks~\cite{canella2019systematic}.
The
latter
-
most
notably
by
the
disclosure
of
Spectre~\cite{Kocher2018spectre}
and
Meltdown~\cite{Lipp2018meltdown}
-
hugely
amplified
the
interest
and
awareness
of
cache
side
channels.
The
two
most
common
attack
vectors
are
\textsc{Flush+Reload}~\cite{DBLP:conf/uss/YaromF14} and
\textsc{Prime+Probe}~\cite{DBLP:journals/joc/TromerOS10,Osvik-2006-CacheAttacksandCo,Liu-2015-Last-LevelCacheSid}.
\textsc{Flush+Reload} relies on shared memory between the attacker and
the
victim
as
well
as
the
\texttt{clflush}
instruction.
This
instruction
takes
a
memory
address
as
a
parameter
and
flushes
the
corresponding
data
from
all
cache
levels.
If
no
data
was
cached
for
that
address,
the
instruction
has
no
effect.
\textsc{Prime+Probe}
on
the
other
hand
does
not
rely
on
shared
memory
or
the
\texttt{clflush}
instruction.
Instead,
the
attack
makes
use
of
eviction
sets
to
flush
the
victim
entry
from
the
cache.
An
eviction
set
is
a
set
of
$w$
addresses
that
map
to
the
same
cache
set,
where
$w$
equals
the
associativity
of
the
cache.
Any
entries
that
are
stored
in
that
cache
set
prior
to
accessing
the
eviction
set
will
be
replaced
by
the
eviction
set.
Since
the
eviction
set
addresses
then
occupy
all
entries
of
the
set,
the
attacker
can
trigger
the
victim
process
and
measure
if
the
victim
accessed
that
set
by
probing
the
eviction
set
addresses
for
a
cache
miss.
If
a
cache
miss
occurs
during
the
probing
phase,
the
attacker
learns
that
the
victim
accessed
the
cache
set.
Since
the
attacker
does
not
have
full
control
over
the
physical
address,
they
can
only
partially
control
the
set
bits
of
the
address,
namely
those
that
overlap
with
the
page
offset
of
the
virtual
address.
However,
the
attacker
can
choose
a
large
initial
set
of
addresses
that
acts
as
an
eviction
set
by
sheer
size,
and
then
reduce
this
set
to
a
minimal
eviction
set
using
algorithms
proposed
in~\cite{vila2019theory,song2019dynamically}.

\paragraph{Cache Covert Channels.}
There
are
a
large
number
of
possibilities
to
transmit
data
from
one
process
to
another
without
an
observer
noticing.
However,
the
timing
behavior
of
caches
is
used
disproportionately
often
in
the
context
of
microarchitectural
attacks
and
covert
channels,
since
it
allows
fast
and
fine-grained
transmission.
Often,
\textsc{Flush+Reload}~\cite{DBLP:conf/uss/YaromF14} is used
for
covert
channels.
Therefore,
the
receiver
first
makes
sure
that
the
shared
address
between
sender
and
receiver
is
not
cached
using
\texttt{clflush}.
Note,
that
this
shared
address
may
be
read-only.
Next,
the
sender
encodes
one
bit
of
the
message
by
either
accessing
the
shared
address
or
not.
The
receiver
then
measures
the
latency
for
an
access
to
the
shared
address.
Only
if
the
access
results
in
a
cache
hit,
the
sender
accessed
the
address.
The
used
side
channel
is
interchangeable
for
any
other
side
channel,
e.g.,
\textsc{Flush+Flush}~\cite{DBLP:conf/dimva/GrussMWM16}
or
\textsc{Prime+Probe}.

This
process
requires
synchronization
between
sender
and
receiver
which
is
not
trivial.
Usually,
each
symbol
is
repeated
for
a
fixed
timeframe
and
the
sender
and
receiver
perform
their
actions
asynchronously.
Due
to
the
repetition,
the
average
latency
will
reveal
whether
a
zero
or
a
one
was
transmitted.
In
order
to
decode
the
incoming
data
stream,
often
self-clocking
signals
like
Manchester-Encoding
are
used~\cite{tanenbaum1996computer}.

\paragraph{Randomized Caches.}
In
an
effort
to
prevent
the
efficient
construction
of
eviction
sets,
a
variety
of
randomized
cache
architectures
have
been
proposed~\cite{qureshi2018ceaser,DBLP:conf/uss/WernerUG0GM19,tan2020phantomcache}.
These
schemes
randomize
the
address-to-cache-set
mapping,
such
that
the
attacker
cannot
easily
construct
eviction
sets,
even
if
they
have
full
control
over
the
physical
address.
One
physical
address
can
map
to
different
indices
in
different
cache
ways.
This
allows
addresses
to
partially
collide
in
one
cache
way
but
not
the
others
and
hence,
weakens
the
properties
of
eviction
sets.
It
has
been
shown
that
finding
\textit{fully
congruent}
eviction
sets
is
not
feasible
in
reasonable
time~\cite{DBLP:conf/uss/WernerUG0GM19,purnal2021systematic},
i.e.,
it
is
not
feasible
to
obtain
sets
of
addresses
that
collide
with
the
victim
address
in
\textit{every}
cache
way.
Purnal~\etal\ generalize
the
design
proposals
of
randomized
caches
and
present
the
\textsc{Prime+Prune+Probe}
attack
which
is
a
generic
attack
on
randomized
caches
based
on
probabilistic
eviction
sets~\cite{purnal2021systematic}.
Probabilistic
eviction
sets
contain
addresses
that
are
known
to
collide
in
at
least
one
cache
way
with
the
victim
address.
If
the
probabilistic
eviction
set
contains
enough
of
such
addresses,
the
attacker
has
a
high
probability
of
occupying
all
possible
entries
of
the
victim
address.
By
changing
the
randomization
function
frequently,
attacks
based
on
\textsc{Prime+Prune+Probe}
can
be
prevented,
albeit
with
some
performance
overhead.
More
recent
proposals~\cite{DBLP:conf/uss/SaileshwarQ21,DBLP:journals/corr/abs-2104-11469}
combine
randomization
with
further
measures
to
prevent
\textsc{Prime+Prune+Probe}
attacks
by
design.
Both
schemes
aim
to
hide
the
effects
of
victim
cache
accesses
by
freeing
entries
in
the
cache
before
conflicts
occur.

\subsection{The x86 Microarchitecture}
We
now
discuss
some
microarchitectural
aspects
of
recent
x86
processors.
Thereby,
we
focus
on
Intel
processors
although
the
general
information
holds
for
AMD
processors
as
well.
Since
most
of
the
internals
of
these
processors
are
not
public,
we
rely
on
prior
reverse
engineering
efforts
and
the
sparse
public
documentation.

\paragraph{Store Architecture.}
Every
fetched
instruction
is
converted
from
the
visible
x86
instruction
to
one
or
more
$\mu
OPs$
and
is
inserted
to
the
pipeline.
Once
a
write
$\mu
OP$
is
executed,
the
write
is
forwarded
to
the
\textit{store
buffer}
(SB).
On
the
Skylake
microarchitecture,
the
SB
can
hold
up
to
56
entries~\cite{Mandelblat2015}.
Then,
the
L1
cache
is
queried.
If
the
request
results
in
a
cache
hit
and
the
respective
cache
line
is
in
modified
or
exclusive
state
(i.e.
the
line
is
owned
by
the
cache),
the
data
will
be
written
into
the
L1
cache.
Otherwise,
a
\textit{request
for
ownership}
(RFO)
is
issued
and
a
\textit{line
fill
buffer}
(LFB)
is
allocated
to
track
the
outstanding
write.
On
Sandy
Bridge
processors,
there
are
10
LFBs~\cite{Intel2016b}
although
unofficial
sources
report
12
LFBs
for
more
recent
CPU
generations.
According
to
the
documentation,
the
SB
entry
remains
active
until
after
the
store
instruction
retires,
i.e.
the
SB
entry
only
retires
after
the
L1
cache
line
is
filled~\cite{Intel2016b}.

\paragraph{Serializing Instructions vs. Ordering Instructions.}
The
x86
ISA
offers
a
set
of
\textit{serializing}
instructions
and
\textit{ordering} instructions that can be used to ensure the intended
order
of
instructions
and
therefore
prevent
unwanted
effects
of
out-of-order
execution
and
speculation~\cite[Sec.
8.3]{Intel2016}.
The
ordering
instructions
are
\texttt{sfence},
\texttt{lfence}
and
\texttt{mfence}
which
are
accessible
from
userspace.
The
store-fence
(\texttt{sfence})
instruction
ensures
that
all
write
instructions
prior
to
the
fence
become
globally
visible
before
those
after
the
fence~\cite[P.
4-599]{Intel2016a}.
The
load-fence
(\texttt{lfence})
does
the
same
for
load
instructions~\cite[P.
3-529]{Intel2016a}
and
the
memory-fence
(\texttt{mfence})
combines
both
fences
to
ensure
that
all
loads
and
stores
before
the
fence
become
globally
visible
before
any
load
or
store
after
the
fence~\cite[P.
4-22]{Intel2016a}.

Opposed
to
these
memory-ordering
instructions,
serializing
instructions
enforce
all
modifications
on
the
processor
state
made
by
any
instruction
before
the
serialization
must
be
completed
before
the
next
instruction
is
fetched.
This
poses
a
very
strong
serialization
since
new
instructions
can
only
enter
the
pipeline
after
all
prior
tasks
are
finished.
Importantly,
serializing
instructions
also
drain
the
SB
with
any
outstanding
write
operations
before
the
next
instructions
are
fetched.
On
Intel
processors
there
are
three
non-privileged
serializing
instructions,
namely
\texttt{cpuid},
\texttt{iret} and \texttt{rsm}~\cite[Sec. 8.3]{Intel2016}. While the two
latter
perform
actions
that
would
cause
significant
side-effects
for
the
following
program
execution,
\texttt{cpuid} only affects the values of the registers \texttt{eax},
\texttt{ebx}, \texttt{ecx} and \texttt{edx}. This makes it a formidable
candidate
to
serialize
instructions
in
any
non-privileged
program.
According
to
the
AMD
documentation,
on
AMD
processors
the
\texttt{mfence}
instruction
is
also
a
full
serializing
instruction~\cite[P.
206]{AMD2021}.
\section{Observations on Write-After-Write}
\label{write+write:sec:write_channel}
In
this
section
we
first
provide
details
on
the
\ww{}
side
channel.
We
give
a
brief
summary
on
the
side
channel,
reverse
engineer
the
exact
collision
criteria
and
reason
about
the
origins
of
the
side
channel
leakage.
We
then
take
a
look
at
the
channel
noise
which
yields
our
second
observation,
namely
the
clock
pattern
in
the
write
latency.
Finally,
we
discuss
the
findings
and
identify
affected
CPUs.

\subsection{\ww{} Side Channel}
The
\ww{}
side
channel
exploits
differences
in
the
timing
behavior
of
two
write
operations
based
on
features
of
the
physical
address.
In
a
nutshell,
we
observe
that
if
a
write
operation
is
issued
to
a
given
address,
a
subsequent
write
to
\textit{some}
addresses
is
slower
than
a
subsequent
write
to
\textit{some
other}
addresses.
We
found
in
particular,
that
if
the
physical
address
of
the
first
and
the
second
write
share
some
of
the
lower
address
bits,
the
second
write
will
be
slower
than
if
they
do
not
share
those
bits.
We
reverse
engineer
the
exact
bits
of
the
address
matching
function
in
\autoref{write+write:sec:reversing}.
In
the
following,
we
refer
to
addresses
that
match
by
this
function
as
\textit{colliding addresses}.

\begin{figure}
\centering
\definecolor{black_coral}{HTML}{5b616a}

\begin{tikzpicture}
\begin{axis}[
	width = 3in,
	height = 4cm,
	axis line style = thick,
	axis lines=left, 
	bar width=0.1cm,
	ymin=0,
	xlabel={Clock Ticks},
	ylabel style={align=center},
	ylabel = Number of Cases,
	legend style={at={(0.5,-0.4)},anchor=north, legend columns=-1, draw=none}
]

\addplot[blue!80!white, dashed, ultra thick,fill=blue,fill opacity=.1] coordinates
{(1254,0) (1256,1) (1258,1) (1260,0) (1262,0) (1264,1) (1266,5) (1268,7) (1270,13) (1272,12) (1274,10) (1276,12) (1278,6) (1280,4) (1282,28) (1284,92) (1286,238) (1288,501) (1290,579) (1292,512) (1294,394) (1296,251) (1298,119) (1300,31) (1302,11) (1304,10) (1306,2) (1308,8) (1310,2) (1312,4) (1314,2) (1316,3) (1318,3) (1320,4) (1322,0) (1324,1) (1326,4) (1328,3) (1330,8) (1332,4) (1334,5) (1336,3) (1338,3) (1340,2) (1342,5) (1344,3) (1346,1) (1348,3) (1350,0) (1352,1) (1354,0) (1356,0) (1358,1) (1360,0) (1362,1) (1364,2) (1366,2) (1368,0) (1370,0)};
    \addlegendentry{No Collision}
    
\addplot[blue!80!white, ultra thick,fill=blue,fill opacity=.2] coordinates
{(1254, 0) (1256, 0) (1258, 0) (1260, 0) (1262, 0) (1264, 1) (1266, 4) (1268, 2) (1270, 2) (1272, 0) (1274, 0) (1276, 1) (1278, 0) (1280, 3) (1282, 1) (1284, 5) (1286, 14) (1288, 11) (1290, 8) (1292, 8) (1294, 10) (1296, 5) (1298, 42) (1300, 190) (1302, 357) (1304, 441) (1306, 510) (1308, 513) (1310, 338) (1312, 205) (1314, 74) (1316, 31) (1318, 9) (1320, 5) (1322, 4) (1324, 3) (1326, 7) (1328, 5) (1330, 4) (1332, 6) (1334, 4) (1336, 3) (1338, 6) (1340, 2) (1342, 3) (1344, 2) (1346, 3) (1348, 0) (1350, 0) (1352, 4) (1354, 4) (1356, 4) (1358, 4) (1360, 3) (1362, 7) (1364, 3) (1366, 2) (1368, 2) (1370, 1)};
    \addlegendentry{Collision}

\end{axis}
\end{tikzpicture}
\caption{\label{write+write:fig:write+write_hist}Distribution of write times over 3000 iterations for a conflicting address a random address.}
\end{figure}
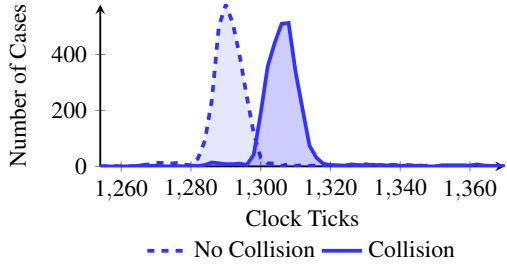

A
minimal
proof-of-concept
pseudocode
is
shown
in
\autoref{write+write:lst:poc_code}.
We
inserted
additional
instructions
that
enforce
the
execution
order
during
the
measurement
and
export
the
timestamp
of
the
\texttt{rdtscp}
instruction.
For
now,
we
set
the
goal
to
find
whether
a
candidate
address
collides
with
a
given
target
address
and
therefore
causes
a
slower
write
access
during
the
measurement.
To
test
this,
the
target
address
is
first
flushed
from
the
cache
using
the
unprivileged
\texttt{clflush}
instruction
of
x86.
The
flushing
can
be
done
at
any
time
during
the
attack
as
long
as
it
is
made
sure
that
the
data
is
not
cached
when
accessed
during
the
measurement.
Then,
a
write
operation
is
issued
to
the
candidate
address.
The
write
is
followed
by
a
\texttt{cpuid}
instruction
which
is
crucial
for
the
success
of
\ww{}.
It
makes
sure
that
the
first
write
instruction
is
retired
before
the
timing
measurement
begins.
Note,
that
\ww{}
does
not
work
if
an
ordering
instruction
like
\texttt{mfence}
is
used
instead.
Opposed
to
ordering
instructions,
\texttt{cpuid}
crucially
also
drains
the
internal
store
buffer.
In
the
final
step,
the
latency
of
a
write
operation
to
the
uncached
target
address
is
measured.
The
distribution
of
the
measured
write
latency
for
an
address
that
is
known
to
collide
with
the
target
(solid)
and
one
that
is
known
to
not
collide
with
the
target
(dashed)
on
an
Intel
Xeon
E-2224G
(Coffee
Lake)
is
shown
in
\autoref{write+write:fig:write+write_hist}. Therefore, we repeatedly measured
the
write
latency
to
the
target
address
with
a
random
candidate
address
\textit{directly followed} by the measurement with a candidate that collides
with
the
target.
From
the
figure
it
is
clear
that
the
distributions
can
be
distinguished
easily.

\begin{lstlisting}[language={[x64]Assembler},caption=AT\&T syntaxed pseudo-code assembly for the \ww{} PoC., label=write+write:lst:poc_code]
CLFLUSH ([target])
MOVQ rax, ([candidate])
CPUID
RDTSCP
MOVQ rdx, ([target])
RDTSCP
\end{lstlisting}

\subsection{Collision Criteria}
\label{write+write:sec:reversing}
\begin{figure}
\centering
\includegraphics[width=.8\linewidth]{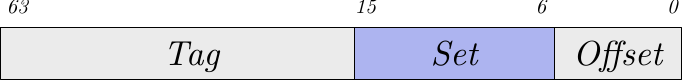}
\caption{\label{write+write:fig:colliding_bits}The bits used for the lower address match are highlighted in blue. The address is divided into the L3 cache addressing parts.}
\end{figure}

We
now
focus
on
reverse
engineering
the
criteria
under
which
two
addresses
collide,
and
therefore
influence
the
write
latency
of
each
other.
For
this,
we
again
use
the
Coffee
Lake
Intel
Xeon
E-2224G,
however
we
later
verified
that
the
observations
hold
for
all
tested
Intel
CPUs,
listed
in
\autoref{write+write:tab:affected_cpus}.
For
the
reverse
engineering,
we
ran
tests
where
we
fix
the
target
address
and
perform
\ww{}
on
each
address
of
a
large
array
to
find
those
colliding.
We
gather
those
addresses
that
led
to
an
increased
latency
for
further
analysis.
Using
the
\textit{libtea}
framework~\cite{easdon2022rapid},
we
analyzed
several
properties
of
the
addresses
and
found
that
the
physical
address
of
each
analyzed
address
matches
the
target
in
the
10-bit
range
between
bit
6
and
15
as
shown
in
\autoref{write+write:fig:colliding_bits}.
We
verified
this
by
allocating
and
testing
possible
physical
addresses
that
only
differ
in
the
bit-range
of
interest
and
found
that
none
of
these
candidates
influenced
each
other.
This
rules
out
the
possibility
that
the
function
combines
some
parts
using
a
more
complex
technique
(as
it
is
for
example
the
case
for
the
cache-slice
selection).
We
repeated
this
process
multiple
times
to
account
for
false
positives
and
the
influence
of
noise.

We
further
attempted
to
mount
\ww{}
across
multiple
processor
cores
and
hyperthreads.
Therefore,
we
tried
a
synchronized
and
a
non-synchronized
variant.
Both
variants
split
the
\ww{}-code
in
two
threads,
one
flushing
the
target
address
and
measuring
the
access
time
to
it,
the
other
repeatedly
writing
to
the
candidate
address.
The
synchronized
variant
utilizes
mutexes
to
ensure
the
correct
order
of
instructions,
the
non-synchronized
variant
performs
the
operations
in
a
loop
without
synchronization.
We
did
not
find
clear
indications
that
\ww{}
can
be
exploited
across
hyperthreads
or
CPU
cores.
We
therefore
conclude
that
either
the
addressing
function
implements
some
additional
context-awareness
(e.g.,
by
matching
the
id
of
the
originating
core),
the
noise
level
makes
it
immensely
hard
to
observe
the
effect,
or
the
observed
hardware
structure
is
not
shared
among
cores
/
hyperthreads.
The
load-
and
store
buffers
are
believed
to
be
partitioned
in
recent
CPUs~\cite{Kanter2012}.
Hence,
it
is
likely
that
the
hardware
that
causes
\ww{}-leakage
is
also
partitioned.

The
measurable
timing
difference
of
\ww\ is
either
an
artifact
of
a
false
dependency
check
within
the
CPU
core,
or
a
conflicting
use
of
some
hardware
resource
that
processes
the
write
instructions.
All
our
tested
CPUs
use
exactly
the
10
Bits
identified
with
\ww\ for
L2
and
L3
cache
indexing.
Hence,
we
suspect
that
the
simultaneous
write
access
to
the
two
addresses
causes
a
collision
in
the
set
addressing
process
which
yields
the
measurable
timing
difference.
To
support
this
hypothesis,
we
attempted
to
swap
the
write
instructions
for
non-temporal
writes,
i.e.,
writes
that
do
not
affect
the
cache.
After
that,
\ww{}
no
longer
works.
We
believe
that
the
process
of
set
allocation
is
similar
in
the
L2
and
L3
cache.
Since
the
store
buffers
are
partitioned
in
recent
Intel
CPUs~\cite{Kanter2012}
we
suspect
that
the
hardware
for
allocating
the
cache
sets
is
also
partitioned
and
the
structural
issue
that
causes
the
measurable
timing
delay
is
to
be
found
within
this
logic.
Since
the
LLC
allows
for
cross-core
attacks,
we
focus
on
the
implications
on
the
LLC
set-contention
in
the
following.
We
show
how
\ww\ can
be
used
for
efficient
LLC
attacks
in
\autoref{write+write:sec:evsets}.

\subsection{Dealing with Noise}
\label{write+write:sec:noise_handling}

\begin{figure}
\centering
\definecolor{black_coral}{HTML}{5b616a}

\begin{tikzpicture}
\begin{axis}[
	width = 3in,
	height =4cm,
	axis line style = thick,
	axis lines=left, 
	bar width=0.1cm,
	ymin=0,
	xlabel={Clock Ticks},
	ylabel style={align=center},
	ylabel = Number of Cases,
	legend style={at={(0.5,-0.4)},anchor=north, legend columns=-1, draw=none}
]
 \addplot[black, dashed, ultra thick] (1358, 8);
 \addlegendentry{No Collision}
 
 \addplot[black, ultra thick] (1358, 8) (1360, 9);
 \addlegendentry{Collision}

\addplot[red!80!white, ultra thick, dashed, fill=red,fill opacity=.2] coordinates
{(1292, 3) (1294, 0) (1296, 1) (1298, 1) (1300, 2) (1302, 3) (1304, 9) (1306, 13) (1308, 11) (1310, 15) (1312, 20) (1314, 21) (1316, 10) (1318, 20) (1320, 20) (1322, 29) (1324, 139) (1326, 255) (1328, 295) (1330, 311) (1332, 413) (1334, 344) (1336, 272) (1338, 131) (1340, 59) (1342, 61) (1344, 91) (1346, 106) (1348, 87) (1350, 72) (1352, 38) (1354, 18) (1356, 11) (1358, 4) (1360, 2) (1362, 3) (1364, 3) (1366, 3) (1368, 2)};

\addplot[red!80!white, ultra thick, fill=red,fill opacity=.1] coordinates
{(1292, 0) (1294, 0) (1296, 0) (1298, 0) (1300, 0) (1302, 0) (1304, 2) (1306, 2) (1308, 3) (1310, 1) (1312, 1) (1314, 4) (1316, 4) (1318, 6) (1320, 13) (1322, 15) (1324, 15) (1326, 29) (1328, 21) (1330, 15) (1332, 10) (1334, 12) (1336, 28) (1338, 80) (1340, 185) (1342, 252) (1344, 451) (1346, 508) (1348, 500) (1350, 361) (1352, 174) (1354, 73) (1356, 42) (1358, 8) (1360, 9) (1362, 3) (1364, 5) (1366, 7) (1368, 3)};

\addplot[blue!80!white, ultra thick, dashed, fill=blue, fill opacity=.2] coordinates
{(1254,0) (1256,1) (1258,1) (1260,0) (1262,0) (1264,1) (1266,5) (1268,7) (1270,13) (1272,12) (1274,10) (1276,12) (1278,6) (1280,4) (1282,28) (1284,92) (1286,238) (1288,501) (1290,579) (1292,512) (1294,394) (1296,251) (1298,119) (1300,31) (1302,11) (1304,10) (1306,2) (1308,8) (1310,2) (1312,4) (1314,2) (1316,3) (1318,3) (1320,4) (1322,0) (1324,1) (1326,4) (1328,3) (1330,8) (1332,4) (1334,5) (1336,3) (1338,3) (1340,2) (1342,5) (1344,3) (1346,1) (1348,3) (1350,0) (1352,1) (1354,0) (1356,0) (1358,1) (1360,0) (1362,1) (1364,2) (1366,2) (1368,0) (1370,0)};
    
\addplot[blue!80!white, ultra thick, fill=blue,fill opacity=.1] coordinates
{(1254, 0) (1256, 0) (1258, 0) (1260, 0) (1262, 0) (1264, 1) (1266, 4) (1268, 2) (1270, 2) (1272, 0) (1274, 0) (1276, 1) (1278, 0) (1280, 3) (1282, 1) (1284, 5) (1286, 14) (1288, 11) (1290, 8) (1292, 8) (1294, 10) (1296, 5) (1298, 42) (1300, 190) (1302, 357) (1304, 441) (1306, 510) (1308, 513) (1310, 338) (1312, 205) (1314, 74) (1316, 31) (1318, 9) (1320, 5) (1322, 4) (1324, 3) (1326, 7) (1328, 5) (1330, 4) (1332, 6) (1334, 4) (1336, 3) (1338, 6) (1340, 2) (1342, 3) (1344, 2) (1346, 3) (1348, 0) (1350, 0) (1352, 4) (1354, 4) (1356, 4) (1358, 4) (1360, 3) (1362, 7) (1364, 3) (1366, 2) (1368, 2) (1370, 1)};

\end{axis}
\end{tikzpicture}
\caption{\label{write+write:fig:write+write_noise_hist}Two non-successive executions of \ww{}. For each execution, one address that is known to collide with the target (solid) is compared to an address that does not collide (dashed).}
\end{figure}
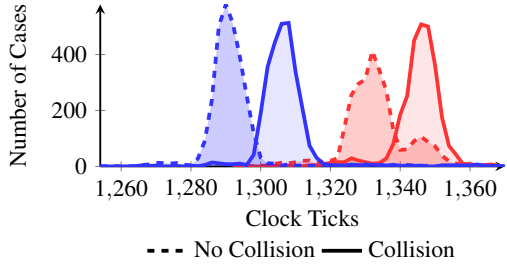

\autoref{write+write:fig:write+write_noise_hist} shows two non-successive
executions
of
the
PoC
code.
For
each
of
these
distributions,
the
target
and
the
candidate
address
are
repeatedly
measured
in
alternating
order.
It
is
clear
that
while
the
distribution
itself
appears
to
be
similar
for
each
run,
the
ideal
threshold
that
distinguishes
colliding
addresses
from
those
that
do
not
collide,
varies
drastically.
As
a
result,
it
is
not
possible
to
make
a
decision
based
on
a
single
measurement
or
even
distribution.
However,
for
measurements
that
are
taken
in
close
succession
like
the
alternating
measurements
that
make
up
the
colliding
and
non-colliding
distribution,
a
distinction
is
simple.
Hence,
the
channel
is
only
stable
over
short
temporal
periods.
This
distinguishes
the
\ww{}
side
channel
from
many
other
CPU
side
channels
like
\textsc{Flush+Reload}
and
\textsc{Prime+Probe}
where
a
threshold
can
be
established
which
reliably
decides
the
two
distributions.
The
reason
for
the
temporal
instability
of
the
channel
can
be
found
in
the
average
write
latency
to
any
address.

\begin{lstlisting}[language={[x64]Assembler},caption=AT\&T syntaxed pseudo-code to measure the write latency., label=write+write:lst:write_latency]
cpuid
rdtscp
movq rdx, ([address])
cpuid
rdtscp
\end{lstlisting}

We
measure
the
write
latency
using
the
code
in
\autoref{write+write:lst:write_latency}
in
a
loop.
The
initial
\texttt{cpuid}
instruction
ensures
no
unfinished
write
instructions
are
in
the
pipeline
at
the
beginning
of
the
measurement.
The
second
\texttt{cpuid}
ensures
that
the
measurement
is
only
stopped
when
the
write
is
completed.

\begin{figure}
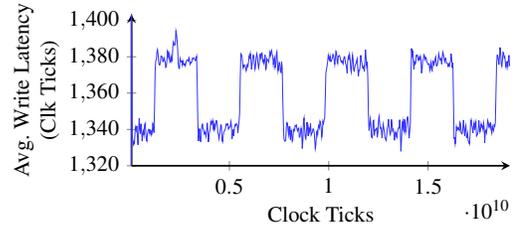

\centering
\includestandalone[width=.8\linewidth]{img/noise_avg}
\caption{\label{write+write:fig:noise_avg} Moving average of the write instruction latency on the Xeon E-2224G.}
\end{figure}

\autoref{write+write:fig:noise_avg}
depicts
the
moving
average
of
the
resulting
latency
measurement.
Surprisingly,
the
graph
represents
a
rather
sharp
clock
signal.
Later
in
this
paper
we
show
how
this
can
be
leveraged
for
cross
core
synchronization.
We
suspect
that
the
observed
behavior
is
an
artifact
of
a
CPU
internal
state-machine.
Although
the
high-
and
low-level
of
the
signal
appear
to
be
stable
in
the
figure,
we
found
that
it
can
slowly
change
over
time
which
might
be
due
to
the
dynamic
frequency
adaption
of
the
CPU.

To
filter
the
noise
and
still
be
able
to
distinguish
colliding
addresses,
it
is
therefore
required
to
take
a
comparative
approach.
In
other
words,
the
results
of
a
measurement
are
only
valuable
in
comparison
to
another
measurement
taken
in
close
succession.
Since
the
addresses
collide
on
10
bits,
the
probability
of
a
collision
for
a
randomly
chosen
address
is
$2^{-10}$.
Hence,
by
choosing
a
random
address
to
compare
the
measurement
against,
the
attacker
has
a
high
probability
of
successfully
gathering
the
addresses
that
collide
with
the
target.
Since
some
of
the
bits
can
even
be
influenced
by
the
virtual
address,
the
attacker
can
also
make
sure
that
the
random
address
does
not
collide
with
the
target.
While
it
is
sufficient
to
test
multiple
iterations
of
accessing
the
target
address
combined
with
the
candidate
address,
directly
followed
by
multiple
iterations
of
accessing
the
target
address
combined
with
the
random
address
(resulting
essentially
in
\autoref{write+write:fig:write+write_hist}),
we
find
that
a
better
way
of
distinguishing
the
two
addresses
is
to
toggle
between
accessing
the
target-
and
the
candidate
address
every
second
iteration.
This
results
in
an
access
pattern
of
\texttt{T-T-C-C-...}
which
avoids
most
of
the
prefetcher
effects
that
are
otherwise
present.
The
measurements
are
summed
for
the
candidate-
and
the
random
address
respectively
such
that
afterwards,
the
mean
latency
of
both
addresses
can
be
computed.
By
subtracting
the
mean
latency
of
the
iterations
with
the
random
address
from
the
mean
latency
of
the
iterations
with
the
candidate
address,
we
can
test
whether
the
distributions
have
a
large
difference
in
their
mean
value
and
hence
conclude
if
the
candidate
collides
with
the
target
address.
If
the
two
addresses
did
not
collide,
then
the
distribution
resulting
from
\ww{}
with
the
random
address
is
similar
to
the
distribution
of
\ww{}
with
the
candidate
address,
resulting
in
a
small
difference
in
means.
We
found
that
a
threshold
of
10
clock
cycles
difference
in
means
after
30
iterations
with
each
address
gives
a
reliable
indication
of
whether
the
two
addresses
collide
on
each
tested
CPU.
The
pseudocode
is
given
in
\autoref{write+write:lst:same_process_c}. It is beneficial to
write
the
code
directly
in
assembly,
using
conditional
moves
instead
of
branch
instructions.
This
prevents
unwanted
effects
from
the
branch
predictor
and
mis-speculation.

\begin{lstlisting}[language={C2},caption=C-flavored pseudo code for same-process testing if two addresses collide with \ww{}., label=write+write:lst:same_process_c]
size_t *random = (size_t*) malloc(8);
bool decision = true;
for(int i=0; i < 2*RUNS; i++;){
decision = iif (decision)
sum_1 += w+w(target, candidate);
else
sum_2 += w+w(target, random);
}
if((sum_1/RUNS)-(sum_2/RUNS) > TH){
// collision
}else{ /* no collision */ }
\end{lstlisting}

\subsection{Discussion}
\label{write+write:sec:discussion}

We
tested
our
implementation
of
the
\ww{}
side
channel
on
various
different
CPUs
which
are
listed
in
\autoref{write+write:tab:affected_cpus}.
All
tested
Intel
CPUs
showed
the
behavior
described
above.
We
therefore
assume
that
most
of
the
recent
Intel
CPUs
will
be
vulnerable
to
Write-After-Write
effects.
We
adapted
\ww{}
to
an
AMD
CPU
but
were
unable
to
identify
similar
effects
and
therefore
have
no
indication
to
believe
that
other
AMD
CPUs
are
affected.
ARM
and
RISC-V
also
feature
serializing
instructions
and
may
therefore
show
similar
behavior
to
the
Write-After-Write
clock
and
could
be
subject
of
further
studies
in
future
work.

\begin{table}
\centering
\caption{\label{write+write:tab:affected_cpus}List of tested CPUs for Write-After-Write effects. }
\begin{tabular}{l |  c c c}
\textbf{CPU} & \textbf{Architecture} &\textbf{W+W} &\textbf{Clk}\\ [0.5ex]
\hline\hline
Intel
Xeon
E-2224G
&
Coffee
Lake
&
\cmark&
\cmark
\\
\hline
Intel
Xeon
W-3223
&
Cascade
Lake
&
\cmark&
\cmark
\\
\hline
Intel
i5-8259U
&
Coffee
Lake
&
\cmark&
\cmark
\\
\hline
Intel
i5-8265U
&
Whiskey
Lake
&
\cmark&
\cmark
\\
\hline
Intel
i7-7600U
&
Kaby
Lake
&
\cmark&
\cmark
\\
\hline
AMD
Ryzen5
5600H
&
Zen3
&
\xmark&
\xmark\\
\end{tabular}
\end{table}

Opposed
to
other
well-known
side
channel
attacks
on
modern
microprocessors
like
\textsc{Flush+Reload}
and
\textsc{Prime+Probe}, the \ww{} channel relies on a  write operation
and
does
not
work
if
an
address
is
only
read.
Furthermore,
it
is
not
possible
to
mount
\ww{}
attacks
across
process
boundaries.
Therefore,
\ww{}
cannot
directly
be
used
to
leak
data
from
other
processes.
However,
in
combination
with
the
aforementioned
side
channels,
\ww{} and the clock synchronization can be useful tools in the
hands
of
an
attacker.
We
show
how
\ww{}
can
be
used
to
construct
eviction
sets
for
traditional
caches
(\autoref{write+write:sec:evsets})
and
on
side-channel
hardened
architectures
like
ScatterCache~\cite{DBLP:conf/uss/WernerUG0GM19}
(\autoref{write+write:sec:evsets_rand}).
Finally,
we
show
how
the
hidden
clock
signal
can
be
used
to
synchronize
covert
channels
(\autoref{write+write:sec:sync}).
\section{Exploiting Write-After-Write}
\label{write+write:sec:exploit}
In
this
section
we
demonstrate
how
\ww{}
can
be
exploited
to
rapidly
create
eviction
sets.
We
start
by
attacking
traditional
caches
on
real
CPUs.
Then
we
use
a
gem5
implementation
of
ScatterCache~\cite{DBLP:conf/uss/WernerUG0GM19}
to
show
how
\ww{} would affect randomized caches. Finally, we use the
Write-After-Write
clock
for
cross-core
synchronization
of
covert
channels.

Unless
stated
otherwise,
all
CPUs
run
an
unmodified
version
of
Ubuntu
20.04.
We
did
not
disable
any
security
/
performance
features
or
isolate
cores.
We
use
the
Intel
Xeon
E-2224G
as
our
main
evaluation
platform.

\subsection{\ww{} for Rapid Cache Attacks}
\label{write+write:sec:evsets}
\textsc{Prime+Probe}~\cite{DBLP:journals/joc/TromerOS10,Osvik-2006-CacheAttacksandCo,Liu-2015-Last-LevelCacheSid} is
one
of
the
most
widely
used
cache
attacks.
Therefore,
the
attacker
needs
to
be
able
to
efficiently
construct
eviction
sets,
allowing
them
to
reliably
observe
accesses
to
the
victim
address.
The
state-of-the
art
algorithms~\cite{song2019dynamically,vila2019theory}
obtain
such
eviction
sets
using
a
top-down
approach.
They
reduce
a
large
set
of
addresses
that
randomly
include
an
eviction
set
and
then
filter
all
addresses
that
are
not
required
for
a
minimal
eviction
set.
Using
\ww{},
it
is
possible
to
construct
eviction
sets
using
a
bottom-up
approach
that
iteratively
adds
addresses
to
the
eviction
set
without
any
privileges.
As
we
will
show,
this
approach
is
much
faster
compared
to
the
top-down
approach.
Moreover,
behavioral
detection
mechanisms
can
likely
be
bypassed
since
the
methodology
is
drastically
different
compared
to
current
algorithms
and
therefore,
the
fingerprint
for
detection
changes.

In
the
following,
we
first
define
the
attacker
model,
then
describe
our
methodology
and
evaluate
the
performance
and
reliability
on
various
processors.

\paragraph{Attacker Model.} The attacker's goal is to create
an
eviction
set
for
a
known
target
address.
We
assume
that
this
address
is
either
directly
accessible,
or
the
attacker
has
access
to
an
address
that
contends
for
the
same
LLC
cache
set
as
the
target
address;
i.e.
the
$i$-bit
after
the
offset
of
the
physical
addresses
match.
If
the
address
is
not
directly
accessible,
the
attacker
can
obtain
a
colliding
address
by
priming
the
cache
and
then
observe
which
address
is
evicted
after
triggering
the
victim
process
(basically
one
iteration
of
\textsc{Prime+Prune+Probe}~\cite{purnal2021systematic}).
We
\textit{do
not}
require
the
attacker
to
know
any
physical
addresses
or
the
mapping
of
virtual
to
physical
addresses.
Furthermore,
we
do
not
make
use
of
huge
pages
which
are
not
always
available.
From
a
microarchitectural
perspective,
we
assume
that
the
CPU
is
vulnerable
to
\ww{}
as
described
in
\autoref{write+write:sec:write_channel}.

\paragraph{Methodology.}
As
described
in
our
analysis
in
\autoref{write+write:sec:reversing},
\ww{}
allows
the
attacker
to
test
whether
two
virtual
addresses
map
to
the
same
cache
set.
This
does
not
inherently
result
in
addresses
that
collide
in
the
LLC
since
the
L2
cache
also
introduces
\ww\ leakage.
Though
the
L2
cache
uses
the
same
index
bits
on
our
evaluation
CPUs,
it
is
not
partitioned
into
slices.
Hence,
with
\ww,
the
attacker
can
identify
physical
addresses
that
collide
in
the
cache
set-index
but
not
necessarily
the
cache
slice.
For
attacks
on
the
LLC,
the
attacker
needs
to
sort
out
addresses
that
do
not
map
to
the
target
slice.
For
CPUs
with
complex
cache
indexing,
an
undocumented
hash
function
is
used
to
map
the
address
to
a
cache
slice.
Only
if
the
slice
and
the
set
/
index
of
two
addresses
collide,
the
two
addresses
can
potentially
evict
each
other.
The
slice
addressing
function
has
been
reverse
engineered
in~\cite{DBLP:conf/dsd/IrazoquiES15,DBLP:conf/raid/MauriceSNHF15,DBLP:conf/ndss/HundWH13}.
Our
approach
does
not
require
any
knowledge
about
this
function.

The
algorithm
to
construct
eviction
sets
using
\ww\ is
shown
in
\autoref{write+write:alg:evsets}.
Therefore,
we
first
allocate
a
sufficiently
large
memory
area.
The
algorithm
takes
as
input
a
pointer
to
the
target
address,
a
pointer
to
the
memory
area
of
size
$mem\_size$,
and
the
number
of
repetitions
for
each
candidate.
Since
some
of
the
cache
set
bits
can
be
directly
controlled
from
the
virtual
address
space,
we
align
the
lower
12
bits
of
the
first
virtual
address
to
be
tested
with
the
lower
12
bits
of
the
victims
virtual
address.
As
discussed
in
\autoref{write+write:sec:noise_handling},
the
results
are
only
meaningful
when
compared
to
another
measurement
in
close
succession.
Previously
we
mentioned
that
the
target
address
can
be
tested
alternating
with
the
candidate
address
and
a
random
address.
In
that
case,
the
timing
difference
is
reliably
measured
if
the
candidate
address
collides.
Therefore,
the
attacker
needs
to
make
sure
that
the
random
address
maps
to
a
different
cache
set
using
some
bits
of
the
virtual
address.
However,
we
found
that
for
performance
reasons,
a
better
approach
is
to
test
two
candidate
addresses
(i.e.
both
are
$2^{12}$
byte
aligned
to
the
target)
in
parallel
and
compare
the
timing
of
these
instead
of
a
random
address.
This
way,
a
timing
difference
will
be
observed
if
one
of
the
addresses
collide
with
the
target
but
neither
if
none
or
both
do.
Since
it
is
relatively
unlikely
that
two
successive
candidate
addresses
map
to
the
same
cache
index,
two
strategies
are
possible:
The
coverage
optimized
variant
aims
to
retrieve
the
most
conflicts
from
a
memory
range.
In
that
case,
the
two
candidate
addresses
are
$base+i*2^{12}$
and
$base+i*2*2^{12}$
and
$i$
is
increased
by
$2^{12}$
in
each
step.
This
way,
each
address
is
tested
twice
which
reduces
the
probability
of
missing
a
conflict.
The
performance
optimized
version
increases
$i$
by
$2*2^{12}$
in
each
iteration
which
does
not
detect
if
both
candidate
addresses
collide
with
the
target
but
instead
doubles
the
execution
speed.
In
the
following,
we
use
the
performance
variant,
as
shown
in
the
algorithm.

To
reduce
the
error
rate,
each
pair
of
addresses
is
tested
multiple
times
and
the
results
are
averaged.
If
the
absolute
difference
in
means
of
the
two
candidate
addresses
is
larger
than
a
threshold,
one
of
the
candidate
addresses
collides
with
the
target.
The
sign
of
the
difference
indicates
which
of
the
candidate
addresses
collides.
When
a
collision
is
observed,
the
address
is
added
to
the
preliminary
eviction
set.
The
attacker
frequently
tests
if
the
eviction
set
is
functional
by
measuring
whether
it
evicts
the
target
address.
As
long
as
the
eviction
set
is
not
functional,
the
attacker
continues
searching
for
\ww{}
collisions.
When
the
preliminary
eviction
set
is
functional,
the
attacker
can
choose
to
remove
false
positives
and
addresses
that
map
to
different
slices
by
removing
one
address
at
a
time
from
the
set
and
testing
whether
the
remaining
addresses
still
form
an
eviction
set.
This
step
is
not
strictly
necessary
since
the
initial
set
is
also
functional
but
depending
on
the
use-case,
it
may
be
important
for
the
attacker
to
obtain
a
minimal
eviction
set.

\begin{algorithm}
\caption{\label{write+write:alg:evsets}\ww -based eviction set construction for traditional caches.}
\begin{algorithmic}
\State \textbf{Input:} $*target$, $*mem$, $mem\_size$, $rep$
\State $ev \gets \varnothing;$
\State $start \gets align(mem, target);$
\For{$i = 0;$  $i < mem\_size;$  $i\mathrel{+}= 2\cdot 0x1000$}
\State $sum_0 \gets 0;$    $sum_1 \gets 0;$
\State $candidate_0 \gets start+i;$
\State $candidate_1 \gets start+i+0x1000;$
\For{$j = 0;$  $j < rep;$  $j\mathrel{+{+}}$}
\State $decision \gets (j\text{ }\&\text{ }0x2) >> 1;$ \If{$decision == 0$}
\State $sum_0 \mathrel{+}= \text{w+w}(target, candidate_0);$
\Else
\State $sum_1 \mathrel{+}= \text{w+w}(target, candidate_1);$
\EndIf
\EndFor
\State $avg_0 \gets sum_0/(rep/2);$
\State $avg_1 \gets sum_1/(rep/2);$
\If{$avg_1-avg_0 > TH$}
\State $ev \gets ev$ $\bigcup$ $candidate_0;$
\ElsIf{$avg_1-avg_0 < -TH$}
\State $ev \gets ev$ $\bigcup$ $candidate_1;$
\EndIf
\If{$test\_evset(ev) == true$}
\State $break;$
\EndIf
\EndFor
\State $ev \gets \text{ reduce}(ev);$ \Comment{Optional.}\\
\Return $ev$
\end{algorithmic}
\end{algorithm}

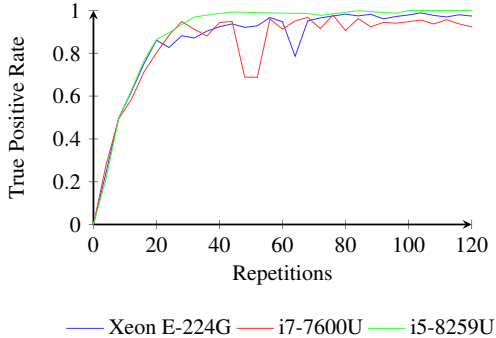
\begin{figure}
\centering
\definecolor{black_coral}{HTML}{5b616a}

\begin{tikzpicture}
\begin{axis}[
	width = 3in,
	height = 5cm,
	axis line style = thick,
	axis lines=left, 
	bar width=0.1cm,
	ymin=0,
	ymax=1,
	xmax=120,
	xlabel={Repetitions},
	ylabel style={align=center},
	ylabel = {True Positive Rate},
	legend style={at={(0.5,-0.4)},anchor=north, legend columns=-1, draw=none}
]

\addplot[blue!80!white] coordinates
{(0, 0)(4, 0.242496)(8, 0.497460)(12, 0.619170)(16, 0.750742)(20, 0.861448)(24, 0.827422)(28, 0.882418)(32, 0.871863)(36, 0.903802)(40, 0.924352)(44, 0.937624)(48, 0.921617)(52, 0.928607)(56, 0.967537)(60, 0.947431)(64, 0.786301)(68, 0.951748)(72, 0.965715)(76, 0.974637)(80, 0.983453)(84, 0.975056)(88, 0.982617)(92, 0.961283)(96, 0.972230)(100, 0.979576)(104, 0.989193)(108, 0.977338)(112, 0.970040)(116, 0.980305)(120, 0.974403)(124, 0.994639)(128, 0.993750)(132, 0.988839)(136, 0.984317)(140, 0.974867)(144, 0.990717)(148, 0.979505)(152, 0.979520)(156, 0.975588)(160, 0.991740)(164, 0.987378)(168, 0.979934)(172, 0.956768)(176, 0.989993)(180, 0.998026)};
\addlegendentry{Xeon E-224G}

\addplot[red!80!white] coordinates
{(0, 0)(4, 0.278984)(8, 0.492818)(12, 0.583857)(16, 0.711864)(20, 0.802619)(24, 0.884778)(28, 0.948565)(32, 0.912628)(36, 0.881295)(40, 0.943828)(44, 0.949002)(48, 0.689230)(52, 0.687830)(56, 0.961043)(60, 0.913079)(64, 0.951579)(68, 0.967954)(72, 0.915439)(76, 0.979121)(80, 0.906808)(84, 0.962655)(88, 0.924123)(92, 0.944453)(96, 0.941035)(100, 0.947993)(104, 0.956001)(108, 0.937264)(112, 0.957151)(116, 0.937761)(120, 0.923727)(124, 0.944506)(128, 0.951542)(132, 0.955474)(136, 0.960032)(140, 0.957282)(144, 0.950186)(148, 0.939402)(152, 0.953728)(156, 0.945950)(160, 0.963059)(164, 0.971497)(168, 0.965025)(172, 0.961061)(176, 0.973469)(180, 0.942839)};
\addlegendentry{i7-7600U}

\addplot[green!80!white] coordinates
{(0, 0)(4, 0.214061)(8, 0.492598)(12, 0.625983)(16, 0.764750)(20, 0.865252)(24, 0.896715)(28, 0.934534)(32, 0.969624)(36, 0.978792)(40, 0.986423)(44, 0.993287)(48, 0.992374)(52, 0.989827)(68, 0.986326)(72, 0.977729)(84, 0.998967)(96, 0.987349)(100, 1.000000)(120, 0.999740)};
\addlegendentry{i5-8259U}

\end{axis}
\end{tikzpicture}
\caption{\label{write+write:fig:evset_success} TPR of a address classification as colliding with the victim address in dependence of the number of repeated measurements. Each data point was averaged over 30 executions.}
\end{figure}

\paragraph{Performance and Reliability.}
In
the
following
we
evaluate
the
\ww{}-based
eviction
set
assembly
on
different
target
CPUs.
The
previously
mentioned
Xeon
W-3223
CPU
is
the
only
one
of
our
test-sample
implements
a
non-inclusive
LLC.
We
therefore
do
not
consider
the
Xeon
W-3223
for
the
eviction
set
use-case.
In
\autoref{write+write:fig:evset_success},
the
confidence
level
of
a
\ww{}-based
observation
after
$2n$
measurements
is
depicted.
This
includes
$n$
measurements
with
the
first
candidate
address
that
are
compared
to
$n$
measurement
to
the
second
candidate
address.
The
True-Positive-Rate
(TPR)
rises
sharply
after
the
first
few
repetitions
and
then
converges
towards
1.
The
characteristic
is
similar
for
all
CPUs
with
the
exception
of
small
outliers.
These
may
be
due
to
scheduler
interruptions
or
system
noise
during
the
measurement.

For
the
construction
of
eviction
sets
it
is
important
that
\textsc{Write+
Write}
reliably
detects
colliding
addresses.
Therefore,
we
now
investigate
the
coverage,
i.e.
how
many
addresses
from
the
memory
area
collide
with
the
target
and
how
many
collisions
are
found
using
\ww{}.
We
use
the
libtea
framework~\cite{easdon2022rapid}
to
compare
the
addresses
that
map
to
the
same
cache
set
to
those
returned
by
\ww{}.
Each
measurement
for
\ww{}
is
repeated
30
times
in
order
to
achieve
a
good
TPR.
We
found
that
in
this
configuration,
\ww{}
detects
about
90\%
of
the
colliding
addresses.

\begin{table}
\centering
\caption{\label{write+write:tab:evsets_performance}Performance evaluation of \ww{}-based eviction set construction. The results are averaged over 500 runs. The time includes filtering of false-positives and addresses that map to other slices.}
\begin{tabular}{l | c c c}
&
\textbf{This
Work}
&
\textbf{Song
\etal~\cite{song2019dynamically}}
\\
[0.5ex]
\hline\hline
Xeon
E-2224G
&
26.6
ms
(96\%)
&
156
ms
(60\%)
\\
\hline
i5-8259U
&
40.6
ms
(86\%)
&
146
ms
(71\%)
\\
\hline
i5-8265U
&
72.5
ms
(81\%)
&
228
ms
(59\%)
\\
\hline
i7-7600U
&
21.1
ms
(95\%)
&
133
ms
(68\%)
\\
\end{tabular}
\end{table}

\autoref{write+write:tab:evsets_performance} shows the performance
for
eviction
set
construction
of
all
tested
CPUs
including
the
reduction
to
a
minimal
eviction
set
and
compares
it
to
the
currently
fastest
algorithm
for
eviction
set
construction
by
Song
\etal~\cite{song2019dynamically}.
We
found
that
the
optimal
performance
for
\ww
-based
eviction
set
construction
can
be
achieved
using
a
tradeoff
between
\ww\ repetitions
and
the
amount
of
false-positive
classifications
for
collisions.
As
shown
in
\autoref{write+write:fig:evset_success},
the
TPR
of
a
collision
classification
becomes
very
high
for
more
than
30
repetitions
of
\ww
.
However,
our
experiments
revealed
that
the
runtime
of
the
eviction
set
construction
algorithm
is
minimal
with
about
10
to
15
repetitions.
This
leads
to
more
false
positives
in
the
preliminary
eviction
set
which
increases
the
runtime
of
the
eviction
set
reduction
to
a
minimal
eviction
set
but
reduces
the
time
to
probe
for
conflicts
in
the
first
part
of
the
algorithm.

We
executed
the
code
by
Song
\etal\ on
our
evaluation
CPUs
to
get
a
clearer
picture
of
the
performance
difference\footnote{Our
results
generally
meet
the
numbers
reported
in
their
paper.
However,
due
to
the
vast
configurability,
it
may
be
that
there
are
slightly
more
optimal
configurations.
We
do
not
expect
major
deviations.}.
For
all
tested
CPUs,
the
\ww
-based
eviction
set
construction
outperforms
the
previous
approach
by
a
factor
of
three
to
six.
The
success
rate
is
also
very
high
throughout
all
our
experiments.
Runs
that
have
been
classified
as
failing
mostly
include
only
one
address
that
is
wrongly
classified
as
collision.
In
such
a
case,
the
false
address
could
be
exchanged
for
a
different
colliding
address
without
much
additional
computing
time.

\subsection{Attacks on Randomized Caches}
\label{write+write:sec:evsets_rand}
The
search
for
effective
countermeasures
to
thwart
cache
side
channels
has
peaked
in
a
number
of
cache
architectures
that
randomize
the
cache
index
using
(partial)
address
encryption~\cite{DBLP:journals/corr/abs-2104-11469,DBLP:conf/uss/WernerUG0GM19,tan2020phantomcache,DBLP:conf/uss/SaileshwarQ21}
to
prevent
efficient
eviction
set
generation.
Cache
randomization
is
generally
considered
to
make
attacks
more
difficult~\cite{DBLP:conf/sp/SongLXLWL21},
even
though
attacks
like
\textsc{Prime+Prune+Probe}~\cite{purnal2021systematic,9251961}
are
still
feasible
on
pure
index-randomization
schemes.
More
recent
designs
try
to
encounter
such
attacks
with
further
security
mechanisms~\cite{DBLP:journals/corr/abs-2104-11469,DBLP:conf/uss/SaileshwarQ21}.
Hence,
we
believe
to
see
some
form
of
index
randomization
to
be
adapted
by
major
CPU
vendors
in
the
not-too-distant
future.

Since
the
principle
of
contention
is
still
present
in
randomized
caches,
i.e.
two
addresses
\textit{can}
still
collide
in
the
cache,
it
is
likely
that
the
implementation
of
the
entry
selection
remains
similar
and
hence,
\ww{}-like
leakage
may
still
exist.

\paragraph{Methodology.}
To
demonstrate
the
threats
of
\ww{}
leakage
in
the
randomized
cache
setting,
we
implement
the
\textsc{Write+
Write}
behavior
in
the
CPU
simulator
gem5~\cite{DBLP:journals/corr/abs-2007-03152}.
On
traditional
caches,
\ww{} causes an increased write latency when two addresses map to the
same
cache
index.
In
randomized
caches,
addresses
may
have
an
index
collision
in
one
cache
way,
but
not
the
others.
In
our
implementation,
the
increased
latency
occurs
when
two
successive
write
operations
are
issued
and
the
address
of
the
second
write
can
map
to
the
same
index
in
the
cache
way
in
which
the
first
write
is
stored.
This
follows
our
assumption
that
\ww\ is
caused
by
a
conflict
in
the
simultaneous
set
allocation
of
two
write
operations.
The
intention
is
that
the
first
write
is
assigned
to
a
set,
and
then
the
second
write
needs
to
be
placed.
If
a
second
write
can
map
to
the
entry
in
which
the
first
one
was
placed,
the
replacement
policy
needs
to
wait
until
the
first
write
has
updated
the
replacement
data.
We
chose
a
conservative
approach
where
the
other
option
would
be
that
a
timing
difference
is
measurable
if
the
two
addresses
can
collide
in
any
cache
way.
The
effect
on
the
security
on
randomized
caches
would
be
equal,
however,
the
latter
approach
would
accelerate
the
attack
even
further.

For
the
randomized
cache,
we
implement
ScatterCache~\cite{DBLP:conf/uss/WernerUG0GM19}
with
random
replacement
policy
in
gem5.
Thus,
each
address
is
randomized
in
every
cache
way
individually,
yielding
$w$
independent
cache
indices
in
a
$w$-way
cache.
The
attacker
model
is
similar
to
the
one
in
the
non-randomized
setting,
although
the
aim
is
no
longer
to
construct
a
minimal
eviction
set
but
instead
a
probabilistic
one;
see~\cite{purnal2021systematic}.
That
is,
since
constructing
a
minimal
eviction
set
would
require
the
attacker
to
find
$w$
fully
congruent
addresses.
This
has
been
shown
to
be
infeasible~\cite{DBLP:conf/uss/WernerUG0GM19}.

The
algorithm
to
construct
a
probabilistic
eviction
set
using
\ww\ is
similar
to
the
algorithm
in
the
non-randomized
cache
and
shown
in
\autoref{write+write:alg:evsets_rand}.
The
main
difference
is
that
the
candidate
addresses
are
no
longer
aligned
to
the
target
address
since
the
attacker
cannot
influence
the
lower
bits
used
for
set
selection
from
the
virtual
address
space.
This
increases
the
search
space
for
colliding
addresses
significantly.
Each
cache
line
holds
64
byte
of
data,
hence,
the
attacker
needs
to
probe
for
colliding
addresses
in
a
64-byte
stride.
Smaller
offsets
would
result
in
addresses
that
map
to
the
same
cache
line
while
larger
offsets
would
skip
potentially
conflicting
addresses.
A
further
difference
is
that
the
attacker
can
no
longer
deterministically
test
if
the
eviction
set
is
complete.
The
test
needs
to
be
conducted
multiple
times
and
the
attacker
needs
to
determine
whether
the
eviction
set
evicts
the
target
with
the
expected
probability
$p_e$.
The
optional
reduction
to
a
minimal
eviction
set
also
differs
from
the
traditional
algorithm.
Instead
of
removing
one
address
at
a
time
and
probing
if
the
eviction
set
is
still
functional,
the
attacker
can
prime
the
target
multiple
times
and
remove
those
addresses
that
are
never
evicted
by
the
target.

\begin{algorithm}
\caption{\label{write+write:alg:evsets_rand}\ww -based eviction set construction for randomized caches.}
\begin{algorithmic}
\State \textbf{Input:} $*target$, $*mem$, $mem\_size$, $rep$
\State $ev \gets \varnothing;$
\For{$i = 0;$  $i < mem\_size;$  $i\mathrel{+}= 2\cdot 0x40$}
\State $sum_0 \gets 0;$    $sum_1 \gets 0;$
\State $candidate_0 \gets mem+i;$
\State $candidate_1 \gets mem+i+0x40;$
\For{$j = 0;$  $j < rep;$  $j\mathrel{+{+}}$}
\State $decision \gets (j\text{ }\&\text{ }0x2) >> 1;$ \If{$decision == 0$}
\State $sum_0 \mathrel{+}= \text{w+w}(target, candidate_0);$
\Else
\State $sum_1 \mathrel{+}= \text{w+w}(target, candidate_1);$
\EndIf
\EndFor
\State $avg_0 \gets sum_0/(rep/2);$
\State $avg_1 \gets sum_1/(rep/2);$
\If{$avg_1-avg_0 > TH$}
\State $ev \gets ev$ $\bigcup$ $candidate_0;$
\ElsIf{$avg_1-avg_0 < -TH$}
\State $ev \gets ev$ $\bigcup$ $candidate_1;$
\EndIf
\If{$test\_evset\_ratio(ev) \approx p_e$}
\State $break;$
\EndIf
\EndFor
\State $ev \gets \text{ reduce}(ev);$ \Comment{Optional.}\\
\Return $ev$
\end{algorithmic}
\end{algorithm}

\paragraph{Performance Evaluation.}
We
configured
gem5
to
use
the
O3
CPU
model
equipped
with
a
small
64
kB,
2-way
associative
L1
cache
and
a
larger
1
MB,
8
way
associative
L2
cache.
Both
cache
levels
use
our
ScatterCache
implementation.
The
randomization
function
is
a
round
reduced
version
of
PRINCE~\cite{DBLP:conf/asiacrypt/BorghoffCGKKKLNPRRTY12}
with
equal
keys
in
the
L1
and
L2
cache.
This
way,
the
generalized
eviction
set
evicts
entries
from
both
levels
of
cache.
In
practice,
this
reduces
the
complexity
of
cache
randomization,
since
the
address
encryption
only
needs
to
be
performed
once
for
cache
lookups
in
any
cache
level.

We
implement
our
attack
and
execute
it
in
gem5's
syscall
emulation
(SE)
mode.
The
SE
mode
executes
the
binaries
in
an
isolated
environment
without
any
operating
system
or
parallel
processes.
Calls
to
OS
functions
are
handled
by
gem5
directly.
Hence,
in
our
setup
there
is
no
noise
or
disturbance
by
scheduler
decisions
or
parallel
processes.
The
numbers
reported
in
the
following
therefore
represent
the
best-case
for
an
attacker.

Constructing
a
probabilistic
eviction
set
with
$p_e=90\%$
for
an
8-way
cache
requires
about
143
addresses
that
collide
with
the
target
in
at
least
one
cache
way
as
shown
in~\cite{purnal2021systematic}.
We
executed
the
attack
10
times
and
the
average
runtime
was
267
ms.
Furthermore,
we
verified
that
the
correctness
of
the
constructed
eviction
set
by
probing
whether
it
evicts
the
target
with
the
expected
probability.

To
compare
our
attack
to
the
established
\textsc{Prime+Prune+Probe}
attack~\cite{purnal2021systematic},
we
also
implement
this
attack
in
the
same
setup.
\textsc{Prime+
Prune+Probe}
repeatedly
fills
large
parts
of
the
cache
and
accesses
the
prime-set
until
there
are
no
more
evictions
within
this
set.
Then,
the
target
address
is
accessed
which
with
some
probability
evicts
one
of
the
attacker
controlled
addresses.
If
an
eviction
is
observed,
the
address
is
added
to
the
generalized
eviction
set.
We
found
that
priming
50\%
of
the
cache
leads
to
a
reasonable
probability
of
observing
an
eviction
while
keeping
the
amount
of
conflicts
in
the
prune
phase
low.
Using
the
same
settings,
we
executed
the
attack
10
times
and
the
average
runtime
was
2.324
s.
Hence,
the
\ww
-based
attack
outperforms
\textsc{Prime+Prune+Probe}
by
a
factor
of
10
in
this
cache
configuration.
In
practice,
priming
and
pruning
a
large
part
of
the
cache
is
difficult
in
the
presence
of
noise
induced
by
parallel
processes.
Any
such
process
may
cause
an
eviction
of
the
pruned
set
which
leads
to
a
false-positive
observation.
\ww\ is less affected by such noise since only two addresses are
used
to
perform
the
measurement.
Therefore,
we
expect
that
our
attack
would
perform
even
better
in
real-world
scenarios
compared
to
\textsc{Prime+Prune+Probe}.

\subsection{Cross-Core Synchronization}
\label{write+write:sec:sync}
If
an
attacker
wants
to
communicate
over
a
covert
channel,
they
must
make
sure
that
the
sender-
and
receiver
process
are
properly
synchronized.
This
is
not
trivial
since
there
is
no
direct
way
for
the
sender
to
communicate
to
the
receiver
that
the
next
symbol
starts
and
in
many
scenarios,
there
is
no
common
clock
(e.g.
when
using
virtualization).
In
prior
work
this
problem
is
often
evaded
by
using
edge
detection
during
post-processing
of
the
received
signal~\cite{kalmbach2020turbocc}
or
using
self-clocking
encodings
such
as
the
phase
/
Manchester
encoding~\cite{tanenbaum1996computer,DBLP:conf/uss/WuXW12,maurice2017hello}.
However,
the
former
suffers
from
noise
being
classified
as
edges
and
limitations
on
how
short
a
single
symbol
can
be,
while
the
latter
reduces
the
channel
entropy
since
two
bits
are
needed
to
transmit
one
bit
of
information.
Moreover,
if
for
example,
the
\textsc{Flush+Reload} channel shall be used for transmission,
the
sender
and
receiver
have
no
way
of
coordinating
the
flush/reload
and
access
steps.
In
practice,
the
sender
and
receiver
perform
the
flush
and
the
access
in
parallel
which
on
average
leads
to
the
low
reload
latency
on
the
receiver
end.
However,
this
approach
is
not
very
stealthy.
If
the
sender
and
receiver
were
to
share
a
precise
clock
source,
they
could
coordinate
the
flush
and
access
in
a
way
that
each
probe
by
the
receiver
yields
precisely
one
bit
of
message.
This
is
exactly
what
the
Write-After-Write
clock
achieves.

\begin{figure*}[ht]
\begin{minipage}[b]{0.48\linewidth}
	\centering
	\includestandalone[width=.8\linewidth]{img/noise_clk_time}
	\caption*{\label{write+write:fig:noise_clk_times}\ding{182}
Repeatedly
measure
the
write
latency
to
any
address.}
\end{minipage}
\begin{minipage}[b]{0.48\linewidth}
	\centering
	\includestandalone[width=.8\linewidth]{img/noise_clk_time_avg}
	\caption*{\label{write+write:fig:noise_clk_times_avg}\ding{183}
Compute
the
moving
average
of
the
write
latency.}
\end{minipage}
\begin{minipage}[b]{0.48\linewidth}
	\centering
	\includestandalone[width=.8\linewidth]{img/noise_clk_movavg}
	\caption*{\label{write+write:fig:noise_clk}\ding{184}
Edge
detection
in
the
average
write
latency.}
\end{minipage}
\hfill
\begin{minipage}[b]{0.48\linewidth}
	\centering
	\definecolor{black_coral}{HTML}{5b616a}

\begin{tikzpicture}
\begin{axis}[
	width = 3in,
	height = 5cm,
	axis line style = thick,
	axis lines=left, 
	ymax=1.3,
	xlabel={Clock Ticks},
	ylabel style={align=center},
	ylabel = Clock Signal,
	legend style={at={(0.5,-0.4)},anchor=north, legend columns=-1, draw=none}
]

\addplot[blue!80!white] coordinates
{(1646897, 0) (1646898, 1) (4188334685, 1) (4188334686, 0) (6331428061, 0) (6331428062, 1) (8483364247, 1) (8483364248, 0) (10626906793, 0) (10626906794, 1) (12776567673, 1) (12776567674, 0) (14963188825, 0) (14963188826, 1) (17112719017, 1) (17112719018, 0) (19216194713, 0) (19216194714, 1) (21367504463, 1) (21367504464, 0) (23559940427, 0) (23559940428, 1) (25700190903, 1) (25700190904, 0)  };
    \addlegendentry{Core A}
    
\addplot[red!80!white] coordinates
{(2075915, 0) (2075916, 1) (4169316327, 1) (4169316328, 0) (6322904379, 0) (6322904380, 1) (8463305485, 1) (8463305486, 0) (10625726939, 0) (10625726940, 1) (12759319271, 1) (12759319272, 0) (14903268507, 0) (14903268508, 1) (17052944601, 1) (17052944602, 0) (19204316901, 0) (19204316902, 1) (21349282293, 1) (21349282294, 0) (23493911139, 0) (23493911140, 1) (25643742197, 1) (25643742198, 0) };
    \addlegendentry{Core B}

\end{axis}
\end{tikzpicture}
	\caption*{\label{write+write:fig:noise_clk_signal}\ding{185}
Use
a
threshold
to
generate
the
final
clock
signal.}
\end{minipage}
\caption{\label{write+write:fig:clock_signal_process} Illustration of the process to generate a synchronized clock signal between two processes, based on an Intel Xeon E-2224G.}
\end{figure*}

In
the
following,
we
demonstrate
how
changes
in
the
average
write
latency
can
be
leveraged
for
cross-core
synchronization
of
multiple
processes.
We
show
how
covert
channels
can
synchronize
a
sender
and
receiver
process
by
observing
the
average
write
latency
to
an
arbitrary
address.
Importantly,
this
address
does
not
need
to
be
shared
between
the
sender
and
receiver
process.
Due
to
the
synchronization,
the
sender
only
needs
to
send
each
symbol
once
and
the
receiver
only
once
measures
the
access
time
to
read
the
symbol.
This
makes
our
approach
much
more
stealthy
compared
to
previous
techniques.

\paragraph{Attacker Model.}
We
assume
that
the
attacker
controls
two
processes
on
the
target
device.
The
goal
is
to
transmit
a
message
from
one
process
to
the
other
over
a
covert
channel.
The
sender
and
the
receiver
process
execute
in
parallel
but
not
necessarily
on
the
same
physical
CPU
core.
For
covert
channels
that
require
shared
memory
(i.e.
\textsc{Flush+Reload}~\cite{DBLP:conf/uss/YaromF14} and
\textsc{Flush+Flush}~\cite{DBLP:conf/dimva/GrussMWM16}), we assume
that
both
processes
can
access
a
shared
memory
resource.
For
\textsc{Prime+Probe}, this is not required. Furthermore, we
assume
that
both
parties
have
access
to
a
precise
timer.
Should
\texttt{rdtsc} not be available, such timers can easily be constructed
as
shown
by
Schwarz
\etal~\cite{DBLP:conf/fc/SchwarzMGM17}.

\paragraph{Methodology.}
As
shown
earlier
(c.f.
\autoref{write+write:fig:noise_avg}),
the
average
write
latency
to
a
given
address
periodically
switches
between
a
\textit{high}
and
a
\textit{low}
state.
The
resulting
signal
already
resembles
a
very
sharp
clock
signal.
We
verified
that
this
change
in
the
write
latency
is
synchronized
across
multiple
CPU
cores.
The
raw
measurement
is
shown
in
\autoref{write+write:fig:clock_signal_process}
(\ding{182}).
While
the
average
latency
is
reasonably
stable,
the
latency
of
single
write
instructions
can
vary
massively,
making
trivial
classification
to
the
low,
or
the
high
state
infeasible.
Therefore,
we
compute
a
running
average
of
the
write
latency
(\ding{183})
and
then
perform
edge
detection
on
that
data
(\ding{184}).
To
generate
the
clock
signal,
we
hence
instantiate
a
loop
that
measures
the
write
latency
to
a
given
address
constantly.
It
does
not
matter
whether
the
address
is
cached,
as
long
as
it
is
either
always
cached
or
always
not
cached.
A
ring
buffer
is
used
to
compute
a
moving
average.
Moreover,
we
implement
a
second
ring
buffer
that
stores
delta
between
the
measured
time
and
the
current
moving
average.
This
way,
the
average
of
the
second
ring
buffer
is
close
to
zero
if
the
mean
write
latency
is
stable,
but
if
the
average
write
latency
changes
abruptly,
the
mean
over
the
second
ring
buffer
will
peak
briefly.
Using
this
technique,
a
simple
threshold
value
is
sufficient
to
detect
changes
in
the
write
latency
and
therefore,
generate
the
clock
signal.
If
the
average
change
in
write
latency
is
above
a
threshold
(e.g.,
15)
and
the
current
clock
is
low,
the
signal
changes
to
high
and
vice
versa.
As
depicted
in
\autoref{write+write:fig:clock_signal_process}
(\ding{185}),
the
synchronization
is
highly
accurate
and
even
in
the
selected
small
time
frame,
no
visible
error
can
be
observed.
We
introduce
two
metrics
to
evaluate
the
accuracy
of
the
received
clock
signal.
The
cycle-to-cycle
jitter
measures
the
mean
difference
of
clock
periods
of
two
successive
cycles.
It
calculates
as
$J_{cc}=mean(|T_j-T_{j+1}|)
\forall
j$.
To
quantify
the
measurement
error
of
two
processes
observing
the
Write-After-Write
clock,
we
define
the
synchronization
error
as
the
mean
difference
between
the
detection
of
a
clock
edge
of
two
processes.
It
calculates
as
$S_{cc}=mean(|T^{(P0)}_j-T^{(P1)}_j|)\forall
j$.

Until
now,
the
clock
period
is
fixed
by
the
characteristics
of
the
write
latency.
However,
if
the
synchronization
error
is
small,
the
sender
and
receiver
can
split
each
clock
period
in
smaller
chunks
and
hence
increase
the
bandwidth.
Therefore,
both
the
sender
and
receiver
need
to
keep
track
of
the
average
clock
period.
The
edges
of
the
Write-After-Write
clock
serve
as
synchronization
marks
from
which
both
the
sender
and
receiver
separate
the
expected
period
in
$n$
timeframes,
each
of
which
is
used
to
transmit
one
bit
of
message.
In
the
following,
we
use
the
\textsc{Flush+Reload}
covert
channel
to
transmit
a
message
from
the
sender
to
the
receiver
process.
If
a
'1'-bit
is
to
be
transmitted,
the
sender
will
access
the
shared
memory
address
on
the
rising
edge
of
the
Write-After-Write
clock.
If
a
'0'-bit
is
transmitted,
the
sender
\textit{does
not}
access
that
address.
The
receiver
then
measures
the
access
(read)
latency
to
the
shared
address
on
each
falling
clock
edge
and
flushes
the
shared
address
afterwards
to
prepare
for
the
next
symbol.
On
average,
transmitting
one
bit
of
message
therefore
only
requires
0.5
memory
accesses
on
the
sender
side,
and
a
single
memory
access
and
a
cache
flush
instruction
on
the
receiver
side.

\begin{table}
\centering
\caption{\label{write+write:tab:clk_rates}Statistics for various CPUs averaged over 100 iterations. The average jitter and the synchronization error is shown in percent of the mean clock period.}
\begin{tabular}{l |  c c c}
\small
\textbf{CPU} & \textbf{Period (clk)} & \textbf{$J_{CC}/clk$} & \textbf{$S_{CC}/clk$}\\ [0.5ex]
\hline\hline
Xeon
E-2224G
&
$4.295
x
10^9$
&
0.13\%
&
0.02\%\\
\hline
Xeon
W-3223
&
$4.294
x
10^9$
&
0.02\%
&
0.03\%\\
\hline
i5-8259U
&
$4.295
x
10^9$
&
0.09\%
&
0.03\%\\
\hline
i5-8265U
&
$4.587
x
10^9$
&
10.03\%
&
0.02\%
\\
\hline
i7-7600U
&
$4.157
x
10^9$
&
5.1\%
&
0.05\%
\end{tabular}
\end{table}

\paragraph{Performance and Reliability.}
We
first
measure
some
characteristics
of
the
Write-After-Write
clock
on
our
target
CPUs.
Therefore,
we
execute
two
processes
on
each
CPU
that
both
measure
the
write
latency.
The
results
are
shown
in
\autoref{write+write:tab:clk_rates}. The measured clock
period
is
similar
in
all
our
measurements.
For
the
two
Xeon
processors
and
the
i5-9259U,
the
jitter
is
low,
indicating
a
very
stable
clock
period.
However,
on
the
i5-8265U
and
the
i7-7600U
the,
the
jitter
is
much
higher.
On
these
CPUs,
we
experienced
a
large
number
of
outliers
during
the
measurement
which
reduces
the
accuracy
of
the
observed
clock
signal.
However,
the
synchronization
error
is
very
low
on
all
tested
CPUs,
i.e.
both
processes
are
equally
affected
by
the
noise
and
hence,
synchronization
is
still
provided.

We
now
use
the
Write-After-Write
clock
to
synchronize
a
covert
channel
communication
using
\textsc{Flush+Reload}
on
the
Xeon
E2224-G
CPU.
Both
the
sender
and
receiver
process
outsource
the
clock
generation
to
a
separate
thread.
This
way,
we
get
the
highest
possible
sampling
rate
of
the
write
latency
which
improves
the
accuracy
of
the
retrieved
clock
signal.
We
furthermore
schedule
each
process
(sender,
receiver
and
two
clock
threads)
on
different
CPU
cores
to
avoid
heavy
noise
disturbance.
During
a
transmission
two
kinds
of
errors
may
occur:
We
classify
a
flipped
bit
as
a
\textit{transmission
error}
and
a
missing
or
added
bit
as
a
\textit{clock
error}.
Transmission
errors
occur
if
the
covert
channel
is
noisy
or
the
threshold
is
not
optimally
chosen.
Clock
errors
are
an
artifact
of
failed
clock
synchronization.
This
may
happen
if
one
of
the
processes
gets
descheduled
by
the
scheduler
and
therefore
misses
a
clock
edge,
or
if
the
clock
signal
is
disturbed
by
system
noise.
In
our
implementation,
clock
errors
also
occur
if
the
receiver
process
is
stopped
too
late
or
too
early
which
might
lead
to
additional
or
missing
symbols.
Since
such
errors
are
easily
detected
in
the
final
message,
we
do
not
count
them
into
the
error
rates.

To
compare
the
sent
and
the
received
data
and
classify
the
errors,
we
use
the
Needleman-Wunsch
algorithm~\cite{NEEDLEMAN1970443}.
The
algorithm
originates
in
bioinformatics
and
can
be
used
to
identify
matches,
mismatches
and
gaps
in
two
input
vectors.
Mismatches
correspond
to
transmission
errors
and
gaps
correspond
to
synchronization
errors
during
the
transmission.
We
configure
the
match
and
mismatch
scores
to
1
and
-1
respectively.
We
set
the
score
for
gaps
to
-8
to
prevent
false
classifications
as
gaps.
We
do
not
implement
any
error
correction
during
the
transmission
which
could
reduce
the
amount
of
gaps
and
mismatches.

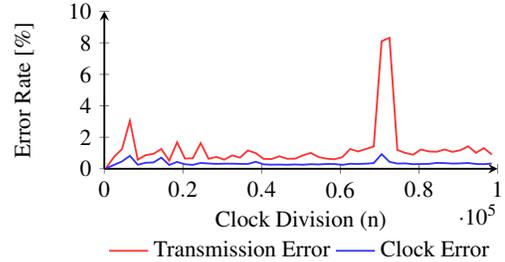
\begin{figure}
\centering
\definecolor{black_coral}{HTML}{5b616a}

\begin{tikzpicture}
\begin{axis}[
	width = 3in,
	height = 4cm,
	axis line style = thick,
	axis y line=left,
	axis x line=bottom,
	bar width=0.1cm,
	ymin=0,
	ymax=10,
	xmin=0,
	xmax=100000,
	xlabel={Clock Division (n)},
	ylabel style={align=center},
	ylabel = {Error Rate [\%]},
	legend style={at={(0.5,-0.4)},anchor=north, legend columns=-1, draw=none}
]

\addplot[red!80!white, thick] coordinates
{(500, 0.06224634637101989) (2500, 0.7650058367035195) (4500, 1.2533957498072124) (6500, 3.0480114797475255) (8500, 0.5719568365780303) (10500, 0.8629214137900632) (12500, 0.9507430635112946) (14500, 1.2544932929124175) (16500, 0.5161041967481683) (18500, 1.6821012959434611) (20500, 0.6466463020392776) (22500, 0.6654221223704571) (24500, 1.6283330104380267) (26500, 0.6337256916924661) (28500, 0.7523530008680612) (30500, 0.5780358611241598) (32500, 0.8512960459739982) (34500, 0.7085969626502264) (36500, 1.162647726063859) (38500, 0.9950187030490998) (40500, 0.6279672214562659) (42500, 0.6157482273914661) (44500, 0.7955862333173247) (46500, 0.6340701542877414) (48500, 0.6405859630998489) (50500, 0.8520813599647403) (52500, 1.0058357273082297) (54500, 0.7400041206725994) (56500, 0.6404939354687889) (58500, 0.6030018806042037) (60500, 0.7336778642034525) (62500, 1.2556724032683209) (64500, 1.100132561786421) (66500, 1.2555793574755225) (68500, 1.4168530947054392) (70500, 8.103843365094566) (72500, 8.319798582343662) (74500, 1.1928614877716404) (76500, 1.0068365599938622) (78500, 0.8952432466094118) (80500, 1.2238442664631421) (82500, 1.0997680459336967) (84500, 1.080342729417605) (86500, 1.224368385018365) (88500, 1.074793539631591) (90500, 1.1745485065486727) (92500, 1.4294500431568713) (94500, 1.0128027741608703) (96500, 1.3122648667348635) (98500, 0.9074546460426214)  };
\addlegendentry{Transmission Error}    

\addplot[blue!80!white, thick] coordinates
{(500, 0.03734594796465274)
(2500, 0.24871139808435316)
(4500, 0.46618755139790835)
(6500, 0.8243955320951386)
(8500, 0.25493879094975114)
(10500, 0.381290042962263)
(12500, 0.40397252395382566)
(14500, 0.7082976331485469)
(16500, 0.24258374049870213)
(18500, 0.4343958634164835)
(20500, 0.28594286459728835)
(22500, 0.2487407586102819)
(24500, 0.3690220482447728)
(26500, 0.3355079333711899)
(28500, 0.31084707189531)
(30500, 0.32319988117559717)
(32500, 0.32325939256772074)
(34500, 0.31067482881420005)
(36500, 0.3046490570518898)
(38500, 0.4469373061258324)
(40500, 0.27972936119773806)
(42500, 0.26102862270013816)
(44500, 0.26728228252940767)
(46500, 0.25482983828159433)
(48500, 0.27363029803764505)
(50500, 0.2549271776475308)
(52500, 0.29201500614848896)
(54500, 0.2735344530145625)
(56500, 0.3046869438257005)
(58500, 0.2983010924691172)
(60500, 0.24870986493459668)
(62500, 0.31702352534581735)
(64500, 0.3044999070485659)
(66500, 0.3230817136167019)
(68500, 0.36051106297539093)
(70500, 0.9267530037795808)
(72500, 0.4412232031336174)
(74500, 0.3294613362910752)
(76500, 0.34167520707386245)
(78500, 0.29191400482775975)
(80500, 0.30443899435466903)
(82500, 0.3106601261901716)
(84500, 0.3725551085459671)
(86500, 0.36031565489328443)
(88500, 0.3293330954813598)
(90500, 0.34197796557297977)
(92500, 0.3668257420010832)
(94500, 0.3044435326650756)
(96500, 0.2921101787995042)
(98500, 0.3292481410377377)};
\addlegendentry{Clock Error}


\end{axis}

\end{tikzpicture}
\caption{\label{write+write:fig:bitrate_error}Transmission errors and clock errors in percent as a function of the clock division rate on the Intel Xeon E-2224G processor.}
\end{figure}

\autoref{write+write:fig:bitrate_error} shows the average
transmission-
and
clock
error
rate
in
percent.
We
therefore
compute
the
average
over
eight
transmissions
of
1kB
data
over
the
covert
channel
using
the
write-synchronization
method.
With
the
exception
of
one
outlier
at
$n\approx
75,000$,
the
error
rates
are
very
low.
The
transmission
error
rate
which
is
purely
influenced
by
the
accuracy
of
\textsc{Flush+Reload}
is
at
about
1\%
while
the
clock
error
rate
is
significantly
smaller
between
0.2\%
to
0.8\%.
The
outlier
may
be
explained
by
a
scheduled
task
that
interrupted
one
of
the
processes.
Since
the
Xeon
E-2224G
CPU
has
only
four
cores,
any
additional
process
directly
competes
for
CPU
time.
We
achieved
a
maximum
transmission
rate
of
up
to
2
kB/s.
Since
the
error
rates
are
still
low
at
this
rate,
we
suspect
that
the
bottleneck
of
the
transmission
speed
lies
in
the
communication
between
the
clock
generating
thread
and
the
sender
/
receiver.
We
also
tested
the
code
on
an
AMD
Ryzen
5
5600H
but
did
not
observe
the
same
clock-like
characteristic
in
the
write
latency.

\section{Mitigation}
\label{write+write:sec:mitigation}

The
mitigation
of
the
Write-After-Write
effects
requires
modification
on
the
microarchitectural
level
by
re-designing
the
aspects
of
the
CPU
that
lead
to
the
behavior
exploited
in
this
work.
This,
however,
requires
significant
change
to
the
design
which
cannot
be
applied
to
currently
deployed
CPUs.
A
less
intrusive
way
to
prevent
Write-After-Write
effects
would
be
a
change
in
the
behavior
of
the
\texttt{cpuid}
instruction.
There
is
no
apparent
reason
why
the
instruction
should
be
unprivileged
and
serializing.
The
fencing
instructions
(\texttt{[s,l,m]fence})
are
well
defined
and
sufficient
to
prevent
speculative
execution
attacks
which
is
a
valid
use-case
for
many
programs.
To
the
best
of
our
knowledge,
beyond
that,
there
is
no
further
scenario
for
an
unprivileged
serializing
instruction.

In-line
with
previous
research
on
cache
side
channels,
making
the
\texttt{clflush} instruction privileged would render \ww{} infeasible
in
unprivileged
environments.
However,
this
does
not
affect
the
clock
synchronization
based
on
the
average
write
latency.
\section{Related Work}
\label{write+write:sec:related}
In
this
section,
we
briefly
summarize
related
work.

\paragraph{Side Channels.}
There
is
a
vast
variety
of
different
side
channels
in
modern
CPUs.
In
the
following,
we
focus
on
memory-related
side
channels
in
desktop-
and
server-grade
CPUs;
for
recent
surveys,
see~\cite{DBLP:journals/csur/LouZJZ21,DBLP:journals/scn/SuZ21,DBLP:journals/jhss/Szefer19}.
Caches
are
the
most
commonly
used
source
of
leakage.
Early
timing-based
attacks
could
be
used
to
recover
cryptographic
keys,
among
others
of
AES~\cite{bernstein2005cache},
DES~\cite{DBLP:conf/ches/TsunooSSSM03},
and
RSA~\cite{DBLP:journals/cn/BrumleyB05}
by
measuring
the
overall
runtime
of
the
program.
Eviction-based
cache
side
channels
increase
the
attack
resolution
since
they
allow
the
attacker
to
trace
single
cache
accesses
by
the
victim.
\textsc{Evict+Time}~\cite{Osvik-2006-CacheAttacksandCo} compares the execution
time
of
a
program
before
and
after
some
cache
entries
were
evicted
to
reconstruct
an
AES
key.
\textsc{Flush+Reload}~\cite{DBLP:conf/uss/YaromF14} flushes a shared cache
line
and
measures
whether
it
will
be
reloaded
by
the
victim.
\textsc{Flush+Flush}~\cite{DBLP:conf/dimva/GrussMWM16} operates similarly but instead
of
measuring
the
reload-latency,
it
measures
the
latency
of
a
second
\texttt{clflush}
instruction.
These
attacks
require
shared
memory
between
the
attacker
and
the
receiver.
\textsc{Prime+Probe}~\cite{DBLP:journals/joc/TromerOS10,Osvik-2006-CacheAttacksandCo}
instead
uses
eviction
sets
to
evict
entries
from
the
cache.
\textsc{Reload+Refresh}~\cite{251542} is a variant of \textsc{Prime+Probe} that
reduces
the
amount
of
cache
misses
and
exploits
the
replacement
policy
of
caches.
Both
attacks
rely
on
eviction
sets~\cite{Liu-2015-Last-LevelCacheSid}
that
reliably
evict
a
target
entry
from
the
cache.
An
algorithm
for
finding
such
eviction
sets
has
been
presented
in~\cite{vila2019theory}
and
improved
in~\cite{song2019dynamically}.
Several
randomization-based
cache
designs
have
been
presented
to
prevent
the
construction
of
eviction
sets~\cite{DBLP:conf/isca/WangL07,DBLP:conf/uss/WernerUG0GM19,tan2020phantomcache,DBLP:conf/uss/SaileshwarQ21,DBLP:journals/corr/abs-2104-11469}.
The
\textsc{Prime+Prune+Probe}
attack~\cite{purnal2021systematic,9251961}
targets
randomized
caches
and
constructs
generalized
eviction
sets,
albeit
much
less
performant
compared
to
regular
caches.
Other
side
channels
in
the
memory
hierarchy
have
been
discovered,
most
notably
on
TLBs~\cite{DBLP:conf/uss/GrasRBG18,DBLP:conf/isca/Deng0S19},
DRAM~\cite{DBLP:conf/uss/PesslGMSM16},
and
the
on-chip
ring
interconnect~\cite{DBLP:conf/uss/PaccagnellaLF21}.
The
group
of
MDS
attacks~\cite{DBLP:conf/ccs/CanellaGGGLMMP019,DBLP:conf/sp/SchaikMOFMRBG19}
exploit
speculative
behavior
in
Intel's
store
buffers.

\paragraph{Covert Channels.}
Cross-VM
covert
channel
communications
have
been
studied
in
real-world
environments
on
AWS
systems
in~\cite{DBLP:conf/ccs/RistenpartTSS09,DBLP:conf/ccs/XuBJJHS11}.
Wu
\etal\ use
a
memory-bus-based
covert
channel
in~\cite{DBLP:conf/uss/WuXW12},
and
Xiao
\etal\ exploit
memory
deduplication
for
covert-communication
\cite{10.1145/2382196.2382318}.
Cache-based
covert
channels
have
been
presented
in~\cite{DBLP:conf/dimva/GrussMWM16,percival2005cache,DBLP:conf/dimva/MauriceNHF15,DBLP:conf/ccs/XuBJJHS11}.
In~\cite{maurice2017hello}
it
has
been
shown
that
cache
covert
channels
can
even
be
used
to
establish
ssh
connections
between
the
communication
partners.

\section{Conclusion}
\label{write+write:sec:conclusion}
We
investigated
the
microarchitectural
peculiarities
of
write
instructions
on
recent
Intel
processors.
We
discovered
\ww{}, a new side channel that leaks set contention in the
cache
architecture.
We
used
\ww{}
for
bottom-up
construction
of
cache
eviction
sets
and
in
doing
so,
broke
current
speed
records
for
eviction
set
construction.
Furthermore,
we
demonstrated
that
attacks
on
randomized
caches
can
be
accelerated
significantly
if
\ww\ leakage
is
present.
Therefore,
we
implemented
ScatterCache
in
gem5
and
benchmarked
our
attack
against
the
recent
\textsc{Prime+Prune+Probe}
attack.
We
found
that
the
\ww
-based
attack
outperforms
current
attacks
by
a
factor
of
10
and
expect
an
even
larger
advantage
in
real-world
implementations.
That
is,
since
the
\ww\ algorithm
for
eviction
set
construction
is
much
less
susceptible
to
noise
by
parallel
processes.

Moreover,
we
developed
a
new
approach
to
synchronize
processes
across
CPU
cores.
The
clock-like
nature
of
the
noise
in
the
write
latency
allows
for
accurate
synchronization
and
therefore
more
stealthy
covert
channel
transmissions.

\section*{Acknowledgments}
Funded
by
the
Deutsche
Forschungsgemeinschaft
(DFG,
German
Research
Foundation)
under
Germany’s
Excellence
Strategy
-
EXC
2092
CASA
-
390781972
and
by
the
DFG
under
the
Priority
Program
SPP
2253
Nano
Security
(Project
RAINCOAT
-
Number:
440059533).
"Any
opinions,
findings,
and
conclusions
or
recommendations
expressed
in
this
material
are
those
of
the
author(s)
and
do
not
necessarily
reflect
the
views
of
DFG
or
other
funding
agencies."

\bibliographystyle{plain}
\bibliography{main}

\appendix

\end{document}